\documentclass[12pt,preprint]{aastex}

\usepackage{amsmath}
\usepackage{natbib}
\usepackage{graphicx}
\usepackage{color}
\usepackage[breaklinks,colorlinks,citecolor=blue]{hyperref}
\usepackage{epsfig}
\usepackage{epstopdf}

\begin{document}
\title{Kerr-Newman-AdS Black Hole In Quintessential Dark Energy}

\author{
  Zhaoyi Xu,\altaffilmark{1,2,3,4}
  Jiancheng Wang,\altaffilmark{1,2,3,4}
 }

\altaffiltext{1}{Yunnan Observatories, Chinese Academy of Sciences, 396 Yangfangwang, Guandu District, Kunming, 650216, P. R. China; {\tt zyxu88@ynao.ac.cn,jcwang@ynao.ac.cn}}
\altaffiltext{2}{University of Chinese Academy of Sciences, Beijing, 100049, P. R. China}
\altaffiltext{3}{Key Laboratory for the Structure and Evolution of Celestial Objects, Chinese Academy of Sciences, 396 Yangfangwang, Guandu District, Kunming, 650216, P. R. China}
\altaffiltext{4}{Center for Astronomical Mega-Science, Chinese Academy of Sciences, 20A Datun Road, Chaoyang District, Beijing, 100012, P. R. China}

\shorttitle{Kerr-Newman-AdS Black Hole In Quintessential Dark Energy}
\shortauthors{Z Y. Xu and J C.  Wang}

\begin{abstract}
Quintessential dark energy with pressure $p$ and density $\rho$ is related by equation of state $p=\omega\rho$ with the state parameter $-1<\omega<-1/3$. The cosmological dark energy influence on black hole spacetime are interesting and important. In this paper, we study the Kerr-Newman-AdS solutions of the Einstein-Maxwell equation in quintessence field around a black hole by Newman-Janis algorithm and complex computations. From the horizon structure equation, we obtain the expression between quintessence parameter $\alpha$ and cosmological constant $\Lambda$ if the black hole exists two cosmological horizon $r_{q}$ and $r_{c}$ when $\omega=-2/3$, the result is different from rotational black hole in quintessence matter situation. Through analysis we find that the black hole charge cannot change the value of $\alpha$. But the black hole spin and cosmological constant are opposite. The black hole spin and cosmological constant make the maximum value of $\alpha$ to become small. The existence of four horizon leads seven types of extremal black holes to constraint the parameter $\alpha$. With the state parameter $\omega$ ranging from $-1$ to $-1/3$, the maximum value of $\alpha$ changes from $\Lambda$ to $1$. When $\omega\rightarrow -1$, the quintessential dark energy likes cosmological constant. The singularity of the black holes is the same with that of Kerr black hole. We also discuss the rotation velocity of the black holes on the equatorial plane for $\omega=-2/3,-1/2$ and $-1/3$. For small value of $\alpha$, the rotation velocity on the equatorial plane is asymptotically flat and it can explain the rotation curves in spiral galaxies.

\end{abstract}


\section{INTRODUCTION}
In recent years, cosmological observations found that the universe is accelerating expansion, demanding the existence of dark energy (\cite{2004ApJ...607..665R,2003RvMP...75..559P,2006IJMPD..15.1753C}). The recent measurements of CMB anisotropy by PLANCK also confirmed this results (\cite{2011ApJS..192...18K}). Cosmological tests indicate that the dark energy accounts for 70$\%$ of energy content in the universe. The state equation of the dark energy is very close to the cosmological constant or vacuum energy. Besides the cosmological constant, an important dark energy model is called quintessence (\cite{2006IJMPD..15.1753C}).

The dark energy content such as the cosmological constant or quintessence changes the spacetime structure of black hole. For the case of the cosmological constant, the asymptotic structure of black hole becomes the asymptotical de Sitter spacetime (\cite{1918AnP...361..401K,1999PhRvD..60d4006S}), in which a cosmological horizon exists. For the black hole in quintessence field, the cosmological horizon also exists (\cite{2003CQGra..20.1187K}).

The importance of cosmological constant in high energy astrophysical objects, such as active galactic nuclei and supermassive black holes, has been discussed (\cite{2005MPLA...20..561S}). The spherically symmetric spacetime influenced by $\Lambda$ term is described by the vacuum Schwarzschild-de Sitter spacetime (SdS) (\cite{1918AnP...361..401K}). When the spacetime metric satisfies the axially symmetric case, the vacuum spacetime is described by Kerr-de Sitter  spacetime (KdS) (\cite{1973blho.conf...57C}). In these spacetimes, the motion of test particles or photons have been discussed by many authors (\cite{2010PhRvD..81d4020H,2005CQGra..22.4391K,2004CQGra..21.4743K,2002PhRvD..65h7301L,2011MPLA...26.2923O,2008PhRvD..77d3004S,2000CQGra..17.4541S,2004PhRvD..69f4001S,1983BAICz..34..129S,1991GReGr..23..507S}). For the spherically symmetric black hole in quintessence field, its spacetime solution has been discussed by \citep{2003CQGra..20.1187K}. The universe accelerating expansion demands the state parameter to be in range $-1<\omega<-1/3$. The recent works generalized this result to Kerr black hole by Janis-Newman algorithm (\cite{2016EPJC...76..222G,2015arXiv151201498T}), and the spacetime metric were studied (\cite{2016EPJP..131..275H,2016Ap&SS.361..269O,2016arXiv160609037S}). Following these works, we generalize Kerr black hole solutions to Kerr-Newman black hole solutions in quintessential dark energy. Following we extend the Kerr-Newman solution to the cosmological constant presented case of quintessential dark energy.

In this paper, we want to seek for Kerr-Newman-AdS solution in the quintessence by Janis-Newman algorithm and complex computations, we also discuss the properties of black hole solution. The outline of the paper is as follows. In section II, we introduce the Reissner-Nordstrom black hole in quintessence matter and derive the Kerr-Newman solution through Janis-Newman algorithm. Later we extend quintessence Kerr-Newman black hole to the case of existing cosmological constant. In section III, we study the horizon structure, stationary limit surfaces and singularity of the black hole in Boyer-Lindquist coordinates. In section IV, we calculate the circular geodesics on the equatorial plane. Summary are drawn in Section V.

\section{KERR-NEWMAN-AdS BLACK HOLE SOLUTION IN QUINTESSENCE}
From spherically symmetric Reissner-Nordstrom black hole metric in the quintessence matter, we use Newman-Janis algorithm to get Kerr-Newman black hole metric around by quintessential dark energy. Because Newman-Janis algorithm don't include cosmological constant, we obtain Kerr-Newman-AdS solution around by quintessential dark energy through direct computations.

\subsection{Reissner-Nordstrom Black Hole in the Quintessence}
For the Reissner-Nordstrom black hole in the quintessence, the line element is expressed by
\begin{equation}
ds^{2}=-f(r)dt^{2}+\dfrac{1}{g(r)}dr^{2}+r^{2}d\Omega^{2},
\label{1}
\end{equation}
where $f(r)$ and $g(r)$ are given by (\cite{2003CQGra..20.1187K})
\begin{equation}
f(r)=g(r)=1-\dfrac{2M}{r}+\dfrac{Q^{2}}{r^{2}}-\dfrac{\alpha}{r^{3\omega+1}}.
\label{2}
\end{equation}
In this spacetime formalism, $M$ is the black hole mass and $\alpha$ is the quintessence parameter that represents the intensity of the quintessence field related to the black hole. The parameter $\omega$ describes the equation of state with $\omega=p/\rho$, where $p$ and $\rho$ are the pressure and energy density of the quintessence respectively, in which $\omega$ will not equal $0, 1/3, -1$ if $-1<\omega<-1/3$ can explain the universe accelerating expansion. Thus we have a general form of Reissner-Nordstrom spacetime metric for the Einstein-Maxwell equation representing charge black hole in quintessential field. The parameter $\omega$ determines the property of spacetime metric. If $-1/3<\omega<0$, the spacetime has the asymptotically flat solution. If $-1<\omega<-1/3$, the spacetime has de Sitter horizon, causing the universe acceleration, and reduces to the Reissner-Nordstrom black hole for the $\alpha=0$.

\subsection{Newman-Janis Algorithm and Kerr-Newman Solution in Quintessence Matter}
Now we derive a Kerr-Newman black hole solution in quintessential field via Newman-Janis algorithm. Following Newman -Janis algorithm (\cite{1965JMP.....6..915N,2015GReGr..47...19E,2010CQGra..27p5008C}) and more general discussion (\cite{2014PhRvD..90f4041A}), we get the coordinate transformation as
\begin{equation}
du=dt-\dfrac{dr}{1-\dfrac{2M}{r}+\dfrac{Q^{2}}{r^{2}}-\dfrac{\alpha}{r^{3\omega+1}}},
\label{4}
\end{equation}
and Equation (\ref{1}) is written as
\begin{equation}
ds^{2}=-(1-\dfrac{2M}{r}+\dfrac{Q^{2}}{r^{2}}-\dfrac{\alpha}{r^{3\omega+1}})du^{2}-2dudr+r^{2}d\Omega^{2}.
\label{5}
\end{equation}
Using the null tetrad, we write the metric matrix as
\begin{equation}
g^{\mu\nu}=-l^{\mu}n^{\nu}-l^{\nu}n^{\mu}+m^{\mu}\overline{m}^{\nu}+m^{\nu}\overline{m}^{\mu},
\label{7}
\end{equation}
where the corresponding components are
\begin{equation}
l^{\mu}=\delta^{\mu}_{r},~~~~
n^{\mu}=\delta^{\mu}_{0}-\dfrac{1}{2}(1-\dfrac{2M}{r}+\dfrac{Q^{2}}{r^{2}}-\dfrac{\alpha}{r^{3\omega+1}})\delta^{\mu}_{r},$$$$
m^{\mu}=\dfrac{1}{\sqrt{2}r}\delta^{\mu}_{\theta}+\dfrac{i}{\sqrt{2}r sin\theta}\delta^{\mu}_{\phi},~~~~
\overline{m}^{\mu}=\dfrac{1}{\sqrt{2}r}\delta^{\mu}_{\theta}-\dfrac{i}{\sqrt{2}r sin\theta}\delta^{\mu}_{\phi}.
\label{8}
\end{equation}
For any point in the spacetime, we choose the tetrad in the following manner: $l^{\mu}$ is the outward null vector tangent to the light cone, and $n^{\mu}$ is the inward null vector. $l^{\mu}$ and $n^{\mu}$ are real vectors. $\overline{m}^{\mu}$ indicates the complex conjugate of $m^{\mu}$, and $m^{\mu}$ is a complex vector. In the null tetrad, they satisfy $l_{\mu}l^{\mu}=n_{\mu}n^{\mu}=m_{\mu}m^{\mu}=0,
l_{\mu}n^{\mu}=-m_{\mu}\overline{m}^{\mu}=1,
l_{\mu}m^{\mu}=n_{\mu}m^{\mu}=0$.
Making the complex coordinate transformations on the $(u,r)$ plane as
$u\longrightarrow u-iacos\theta,
r\longrightarrow r-iacos\theta,$
and following the changes of $f(r)\longrightarrow F(r,a,\theta)$, $g(r)\longrightarrow G(r,a,\theta)$ and $\Sigma^{2}=r^{2}+a^{2}cos^{2}\theta$, we write the null tetrad in new coordinate system as
\begin{equation}
l^{\mu}=\delta^{\mu}_{r},~~~~
n^{\mu}=\sqrt{\dfrac{G}{F}}\delta^{\mu}_{0}-\dfrac{1}{2}F\delta^{\mu}_{r},$$$$
m^{\mu}=\dfrac{1}{\sqrt{2}\Sigma}(\delta^{\mu}_{\theta}+ia sin\theta(\delta^{\mu}_{0}-\delta^{\mu}_{r})+\dfrac{i}{sin\theta}\delta^{\mu}_{\phi}),$$$$
\overline{m}^{\mu}=\dfrac{1}{\sqrt{2}\Sigma}(\delta^{\mu}_{\theta}-ia sin\theta(\delta^{\mu}_{0}-\delta^{\mu}_{r})-\dfrac{i}{sin\theta}\delta^{\mu}_{\phi}).
\label{11}
\end{equation}
Using Equation (\ref{7}), we can get the metric tensor $g^{\mu\nu}$ in Eddington-Finkelstein coordinates.
The covariant components of the metric tenser are given by
\begin{equation}
g_{uu}=-F,~~~~~~g_{\theta\theta}=\Sigma^{2},~~~~~~g_{ur}=g_{ru}=-\sqrt{\dfrac{G}{F}},$$$$
g_{\phi\phi}=sin^{2}\theta(\Sigma^{2}+a^{2}(2\sqrt{\dfrac{F}{G}}-F)sin^{2}\theta),$$$$
g_{u\phi}=g_{\phi u}=a(F-\sqrt{\dfrac{F}{G}})sin^{2}\theta,~~~~~~g_{r\phi}=g_{\phi r}=a sin^{2}\theta\sqrt{\dfrac{F}{G}}.
\label{13}
\end{equation}
Finally, we make the coordinate transformations from the Eddington-Finkelstein coordinates $(u,r,\theta,\phi)$ to the Boyer-Lindquist coordinates $(t,r,\theta,\phi)$ as
\begin{equation}
du=dt+\lambda(r)dr,~~~~d\phi=d\phi+h(r)dr,
\label{14}
\end{equation}
where
\begin{equation}
\lambda(r)=-\dfrac{r^{2}+a^{2}}{r^{2}g(r)+a^{2}},~~~h(r)=-\dfrac{a}{r^{2}g(r)+a^{2}},$$$$
F(r,\theta)=G(r,\theta)=\dfrac{r^{2}g(r)+a^{2}cos^{2}\theta}{\Sigma^{2}}.
\label{15}
\end{equation}
In the Boyer-Lindquist coordinates $(t,r,\theta,\phi)$, the Kerr-Newman metric in the Kiselev quintessence is
\begin{equation}
ds^{2}=-(1-\dfrac{2Mr-Q^{2}+\alpha r^{1-3\omega}}{\Sigma^{2}})dt^{2}+\dfrac{\Sigma^{2}}{\Delta_{r}}dr^{2}-\dfrac{2a sin^{2}\theta(2Mr-Q^{2}+\alpha r^{1-3\omega})}{\Sigma^{2}}d\phi dt+\Sigma^{2}d\theta^{2}$$$$
+sin^{2}\theta (r^{2}+a^{2}+a^{2}sin^{2}\theta\dfrac{2Mr-Q^{2}+\alpha r^{1-3\omega}}{\Sigma^{2}})d\phi^{2},
\label{16}
\end{equation}
where
\begin{equation}
\Delta_{r}=r^{2}-2Mr+a^{2}+Q^{2}-\alpha r^{1-3\omega}.
\label{17}
\end{equation}
Through calculating $R_{\mu\nu}$ and $T_{\mu\nu}$, Azreg-Ainou \citep{2014PhRvD..90f4041A} found that this spacetime metric satisfies Einstein equation. When the quintessence does not exist or $\alpha=0$, the spacetime metric reduces to Kerr-Newman black hole (\cite{1965JMP.....6..915N}). If $Q=0$, the spacetime metric reduces to the rotational situation in the Kiselev quintessence black hole(\cite{2015arXiv151201498T}).

\subsection{Kerr-Newman-AdS Solution in Quintessence Matter}
Now we extend the Kerr-Newman solution to the Kerr-Newman-AdS case of quintessential dark energy. First we rewrite the Kerr-Newman metric in quintessence matter as
\begin{equation}
ds^{2}=\dfrac{\Sigma^{2}}{\Delta_{r}}dr^{2}+\Sigma^{2}d\theta^{2}+\dfrac{sin^{2}\theta}{\Sigma^{2}}(adt-(r^{2}+a^{2})d\phi)^{2}-\dfrac{\Delta_{r}}{\Sigma^{2}}(dt-a sin^{2}d\phi)^{2}.
\label{AdS1}
\end{equation}
Using the formula $G_{\mu\nu}=R_{\mu\nu}-\dfrac{1}{2}Rg_{\mu\nu}$, we deriv the Einstein tenser by Mathematica package RGTC as
\begin{equation}
G_{tt}=\dfrac{2[r^{4}-2r^{3}\rho+a^{2}r^{2}-a^{4}sin^{2}\theta cos^{2}\theta]\rho^{'}}{\Sigma^{6}}-\dfrac{ra^{2}sin^{2}\theta\rho^{''}}{\Sigma^{4}},~~~~
G_{rr}=-\dfrac{2r^{2}\rho^{'}}{\Sigma^{2}\Delta_{r}}$$$$ G_{\theta\theta}=-\dfrac{2a^{2}cos^{2}\theta\rho^{'}}{\Sigma^{2}}-r\rho^{''},~~~~
G_{t\phi}=\dfrac{2a sin^{2}\theta[(r^{2}+a^{2})(a^{2}cos^{2}\theta-r^{2})]\rho^{'}}{\Sigma^{6}}-\dfrac{ra^{2}sin^{2}\theta(r^{2}+a^{2})\rho^{''}}{\Sigma^{4}},$$$$
G_{\phi\phi}=-\dfrac{a^{2}sin^{2}\theta[(r^{2}+a^{2})(a^{2}+(2r^{2}+a^{2})cos2\theta)+2r^{3}sin^{2}\theta\rho)]\rho^{'}}{\Sigma^{6}}-\dfrac{rsin^{2}\theta(r^{2}+a^{2})^{2}\rho^{''}}{\Sigma^{4}},
\label{AdS2}
\end{equation}
where $2\rho=\alpha r^{-3\omega}+2M-\dfrac{Q^{2}}{r}$.
For $Q=0$, these Einstein tensors have been obtained \citep{2015arXiv151201498T}. Using Einstein equation with a cosmological constant and Maxwell equation
\begin{equation}
G_{\mu\nu}=R_{\mu\nu}-\dfrac{1}{2}R g_{\mu\nu}+\Lambda g_{\mu\nu}=8\pi T_{\mu\nu},
\label{AdS3}
\end{equation}
\begin{equation}
F^{\mu\nu}_{;\nu}=0; ~~~ F^{\mu\nu;\alpha}+F^{\nu\alpha;\mu}+F^{\alpha\mu;\nu}=0,
\label{AdS4}
\end{equation}
where $F^{\mu\nu}$ is the Faraday tensor, we obtain the Kerr-Newman-AdS solution in quintessential dark energy.

Considering the cosmological constant, we guess the solution of Einstein-Maxwell equation in quintessence matter given by
\begin{equation}
ds^{2}=\dfrac{\Sigma^{2}}{\Delta_{r}}dr^{2}+\dfrac{\Sigma^{2}}{\Delta_{\theta}}d\theta^{2}+\dfrac{\Delta_{\theta}sin^{2}\theta}{\Sigma^{2}}(a\dfrac{dt}{\Xi}-(r^{2}+a^{2})\dfrac{d\phi}{\Xi})^{2}-\dfrac{\Delta_{r}}{\Sigma^{2}}(\dfrac{dt}{\Xi}-a sin^{2}\dfrac{d\phi}{\Xi})^{2},
\label{AdS5}
\end{equation}
where
\begin{equation}
\Delta_{r}=r^{2}-2Mr+a^{2}+Q^{2}-\dfrac{\Lambda}{3}r^{2}(r^{2}+a^{2})-\alpha r^{1-3\omega},$$$$
\Delta_{\theta}=1+\dfrac{\Lambda}{3}a^{2}cos^{2}\theta~~~~~~~~\Xi=1+\dfrac{\Lambda}{3}a^{2}.
\label{AdS6}
\end{equation}
Calculating by Mathematica package RGTC, we also get the Einstein tenser as
\begin{equation}
G_{tt}=\dfrac{2[r^{4}-2r^{3}\rho+a^{2}r^{2}-a^{4}sin^{2}\theta cos^{2}\theta]\rho^{'}}{\Sigma^{6}}-\dfrac{ra^{2}sin^{2}\theta\rho^{''}}{\Sigma^{4}}+\Lambda\dfrac{a^{2}sin^{2}\theta-\Delta_{r}}{\Sigma^{2}},$$$$
G_{rr}=-\dfrac{2r^{2}\rho^{'}}{\Sigma^{2}\Delta_{r}}+\Lambda\dfrac{\Sigma^{2}}{\Delta_{r}},~~~~G_{\theta\theta}=-\dfrac{2a^{2}cos^{2}\theta\rho^{'}}{\Sigma^{2}}-r\rho^{''}+\Lambda\Sigma^{2}$$$$
G_{t\phi}=\dfrac{2a sin^{2}\theta[(r^{2}+a^{2})(a^{2}cos^{2}\theta-r^{2})]\rho^{'}}{\Sigma^{6}}-\dfrac{ra^{2}sin^{2}\theta(r^{2}+a^{2})\rho^{''}}{\Sigma^{4}}+\Lambda\dfrac{a sin^{2}\theta[\Delta_{r}-r^{2}-a^{2}]}{\Sigma^{2}},$$$$
G_{\phi\phi}=-\dfrac{a^{2}sin^{2}\theta[(r^{2}+a^{2})(a^{2}+(2r^{2}+a^{2})cos2\theta)+2r^{3}sin^{2}\theta\rho)]\rho^{'}}{\Sigma^{6}}-\dfrac{rsin^{2}\theta(r^{2}+a^{2})^{2}\rho^{''}}{\Sigma^{4}}$$$$
+\Lambda\dfrac{sin^{2}\theta[(r^{2}+a^{2})^{2}-a^{2}\Delta_{r}]}{\Sigma^{2}}.
\label{AdS7}
\end{equation}

By calculation, we find that the above metric satisfy the Einstein-Maxwell equation in quintessence matter including the cosmological constant.

\section{KERR-NEWMAN-AdS BLACK HOLE IN QUINTESSENCE}
\subsection{Horizon Structures}

In order to know the properties of the black hole, we calculate the horizon structure of the black hole. From the definition of the horizon
\begin{equation}
g^{rr}=0,
\label{18}
\end{equation}
we find that the horizon satisfies the following equation
\begin{equation}
\Delta_{r}=r^{2}-2Mr+a^{2}+Q^{2}-\dfrac{\Lambda}{3}r^{2}(r^{2}+a^{2})-\alpha r^{1-3\omega}=0,
\label{19}
\end{equation}
which depends on $\alpha,a,Q,\Lambda$ and $\omega$. It is very interesting and different from the Kerr Black Hole. $\alpha,a,Q,\Lambda$ and  $\omega$ will determine the horizon number.

\begin{figure}[htbp]
  \centering
  \includegraphics[scale=0.55]{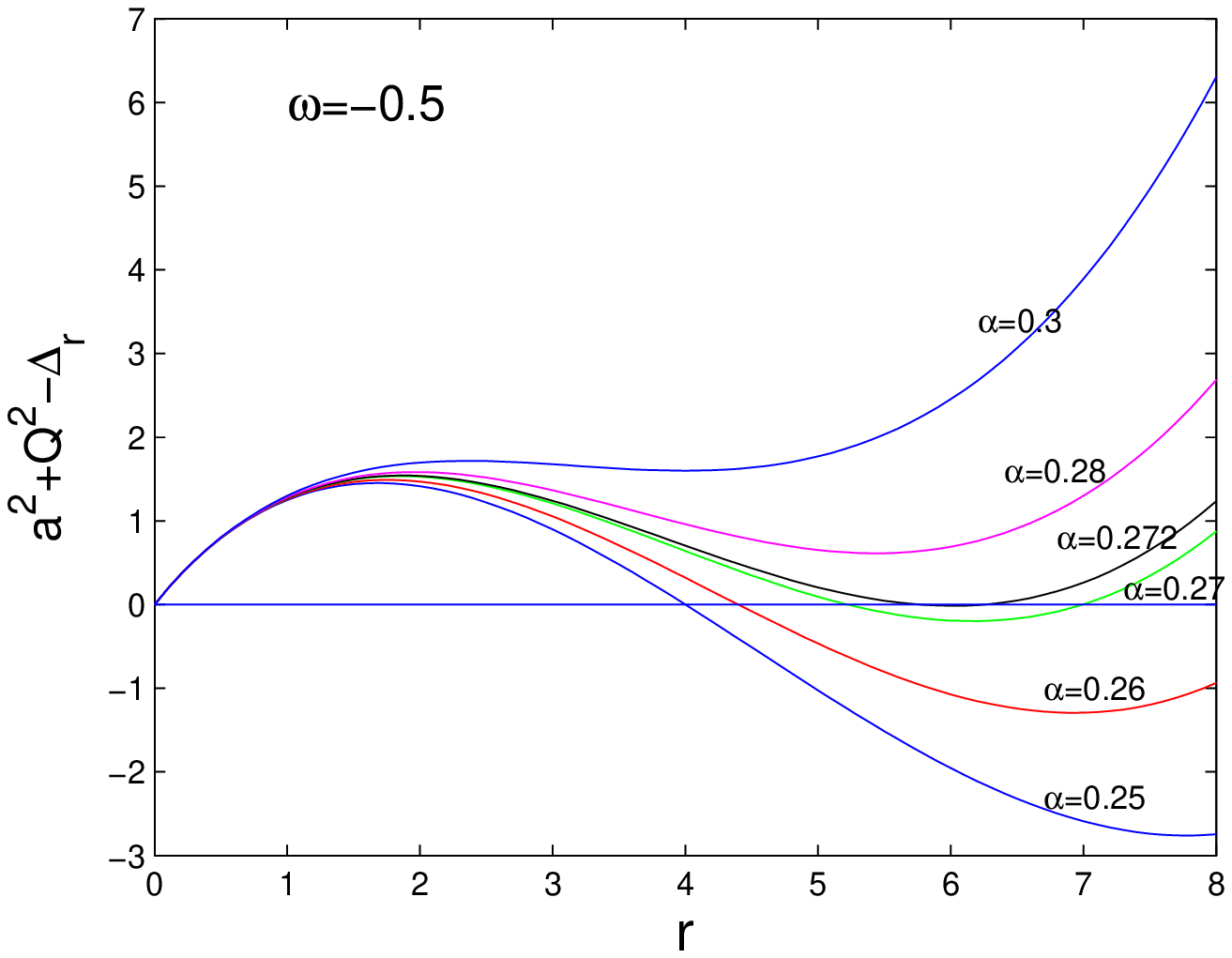}
  \includegraphics[scale=0.55]{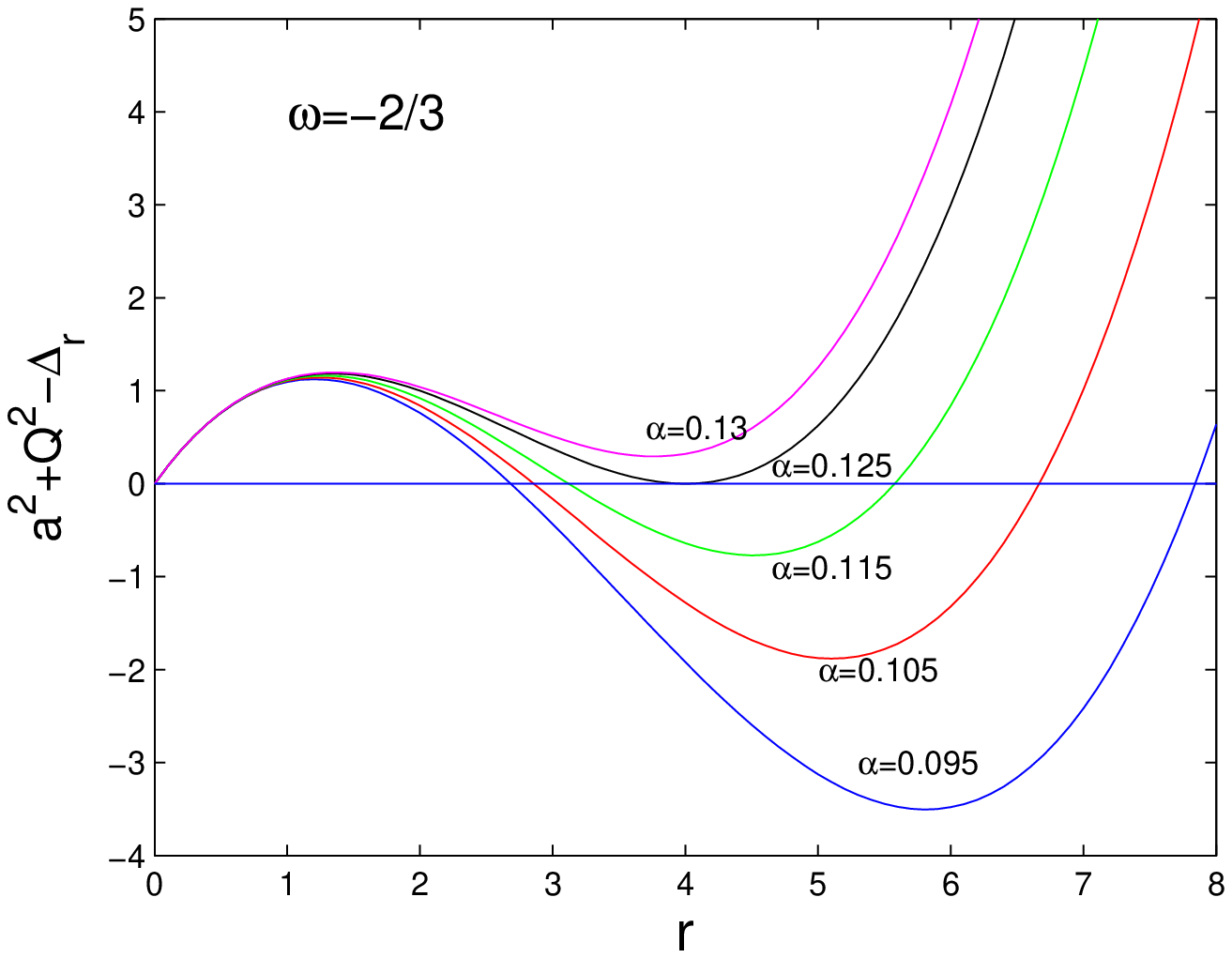}
  \caption{Two pictures show the behavior of $a^{2}+Q^{2}-\Delta_{r}$ with $r$ for fixed $M=1$, in which for different $\omega$, $\alpha$ will satisfy different value when the cosmological horizon exists. Due to the small value of the cosmological constant, there always exist the cosmological horizon $r_{c}$.}
  \label{fig:1}
\end{figure}

It is very convenient to analyse the properties of the black hole if we make (\ref{19}) to become the following form
\begin{equation}
a^{2}+Q^{2}=-r^{2}+2Mr+\alpha r^{1-3\omega}+\dfrac{\Lambda}{3}r^{2}(r^{2}+a^{2}).
\label{20}
\end{equation}
For general $\omega$($-1<\omega<-1/3$) situation, four horizons exist, including Cauchy horizon $r_{in}$ ($r_{-}$), event horizon $r_{out}$($r_{+}$) and two cosmological horizon $r_{q}$ and $r_{c}$, where $r_{q}$ is the cosmological horizon determined by quintessential dark energy and $r_{c}$ is the cosmological horizon determined by the cosmological constant. When the cosmological constant is zero, using the method of \citep{2015arXiv151201498T}, if the cosmological horizon $r_{q}$ exists, we find that the parameter $\alpha$ will satisfy
\begin{equation}
\alpha\leq\dfrac{2}{(1-3\omega)}8^{\omega}.
\label{21}
\end{equation}
For $\omega=-2/3$, we get $\alpha\leq 1/6$ from the equation (\ref{21}), which is the same with the rotational black hole in quintessence matter. For $\omega=-1/2$, we obtain $\alpha\leq\sqrt{2}/5$. These result imply that black hole charge cannot change the value of parameter $\alpha$.

In Figure 1, we show the behavior of $a^{2}+Q^{2}-\Delta_{r}$ with $r$ for fixed $M=1$, in which $\alpha$ satisfy different values for different $\omega$ when the cosmological horizon $r_{q}$ exists. Far away from black hole such as cosmological scale, another cosmological horizon $r_{c}$ exists.

For $\Lambda\neq 0$, the equation exists four roots. If we consider the case of $\omega=-2/3$, the horizon equation becomes
\begin{equation}
r^{4}+\dfrac{3\alpha}{\Lambda}r^{3}+(a^{2}-\dfrac{3}{\Lambda})r^{2}+\dfrac{6M}{\Lambda}r-\dfrac{3}{\Lambda}(a^{2}+Q^{2})=0,
\label{aa1}
\end{equation}
this fourth order algebra equation can be expressed as
\begin{equation}
(r-r_{in})(r-r_{out})(r-r_{q})(r-r_{c})=0.
\label{aa2}
\end{equation}
The existence of cosmological horizon $r_{q}$ will change the parameter $\alpha$. Through analysing the equation, we find that $\alpha$ satisfies
\begin{equation}
(\dfrac{27\alpha^{3}}{64\Lambda^{3}}-\dfrac{3\alpha}{8\Lambda}(\dfrac{a^{2}}{2}-\dfrac{3}{2\Lambda})+\dfrac{3M}{4\Lambda})^{2}+(\dfrac{a^{2}}{6}-\dfrac{1}{2\Lambda}-\dfrac{9\alpha^{2}}{16\Lambda^{2}})^{3}<0.
\label{aa3}
\end{equation}
From above equation, we find that cthe osmological constant make the value of $\alpha$ to become small.

The extremal black hole have seven types. For first type, the inner horizon $r_{-}$ and the outer horizon $r_{+}$ are equal. For second type, the outer horizon $r_{+}$ and the cosmological horizon $r_{q}$ are equal. For third type, the cosmological horizon $r_{q}$ equals to the cosmological horizon $r_{c}$. For fourth type, $r_{in}=r_{out}=r_{q}$. For fifth type, $r_{out}=r_{q}=r_{c}$. For sixth type, $r_{in}=r_{out}=r_{q}=r_{c}$. For seventh type, $r_{in}=r_{out}$ and $r_{q}=r_{c}$. For the different type of extremal black hole, the maximum value of $\alpha$ is also different.

Through analysing these extremal black holes, we find that when $\omega\rightarrow -1$, the quintessential dark energy will like the cosmological constant and $\alpha$ is close to the cosmological constant. For $\omega\rightarrow -1/3$, the cosmological horizon determined by quintessence will be
close to outer horizon and $\alpha$ satisfies $\alpha<1$. From these analysis, we find that the black hole spin and cosmological constant will lead the value of $\alpha$ to become small. With the state parameter $\omega$ ranging from $-1$ to
$-1/3$, the maximum value of $\alpha$ changes from $\Lambda$ to $1$.

\begin{figure}[htbp]
  \centering
  \includegraphics[scale=0.36]{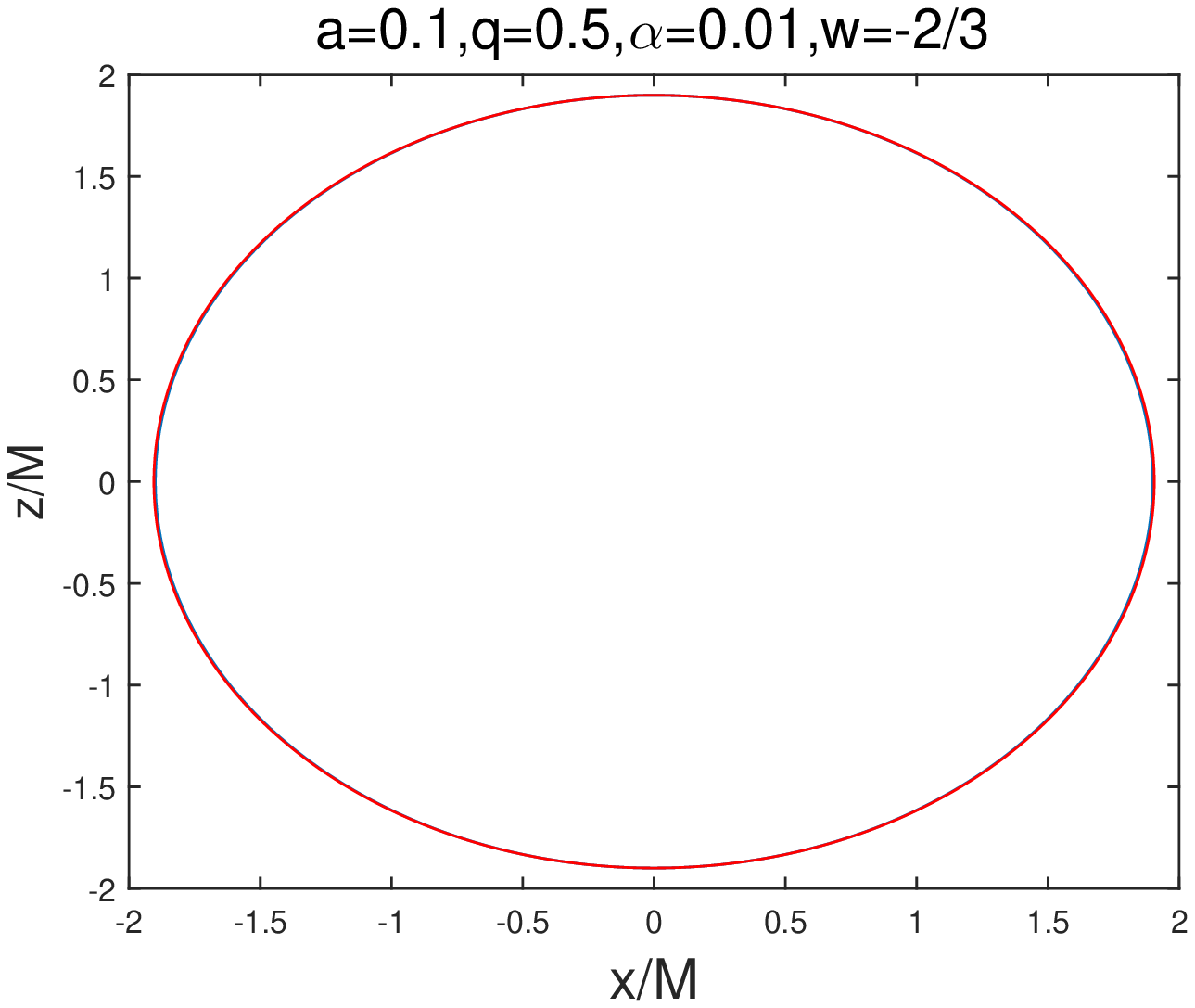}
  \includegraphics[scale=0.36]{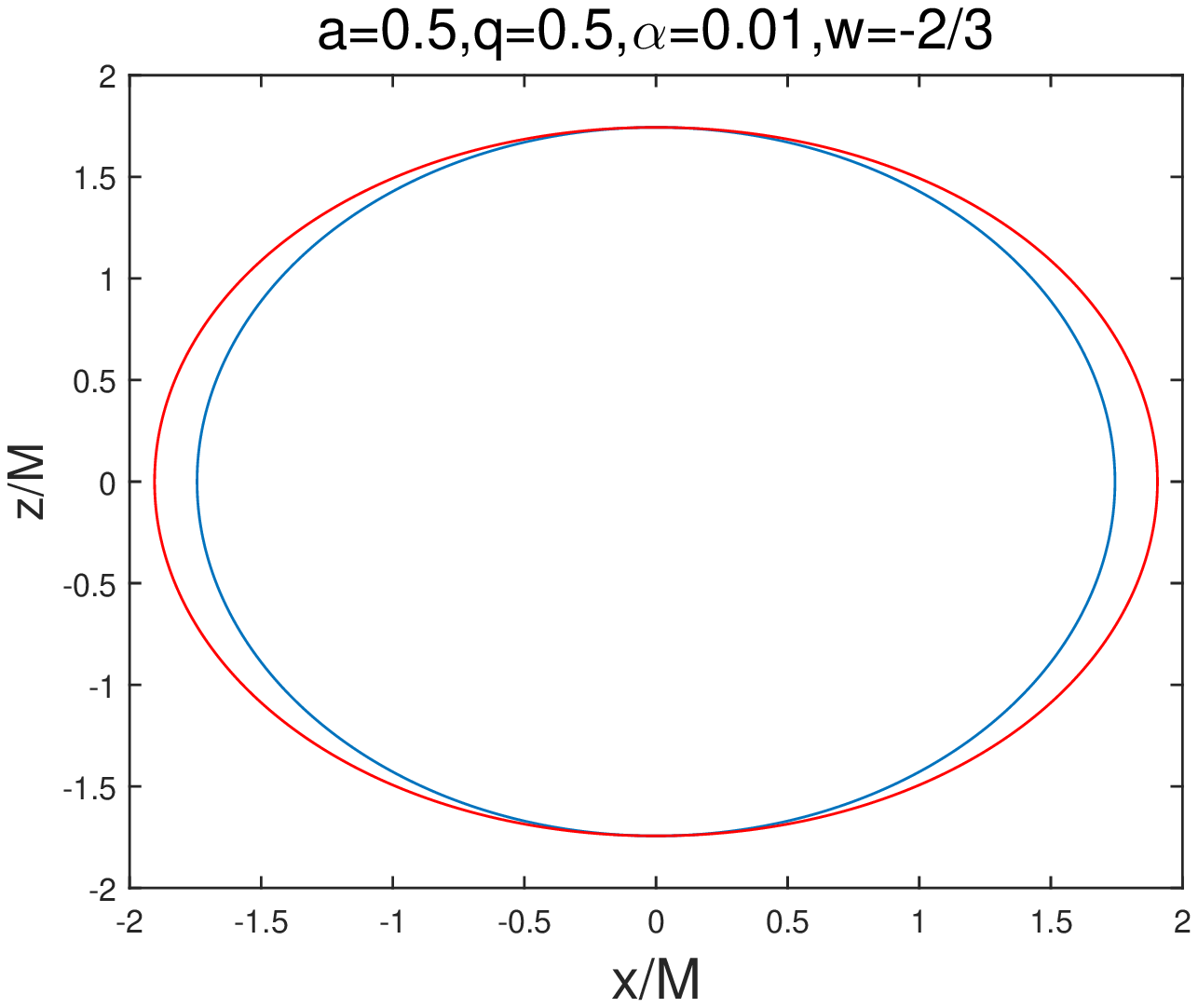}
  \includegraphics[scale=0.36]{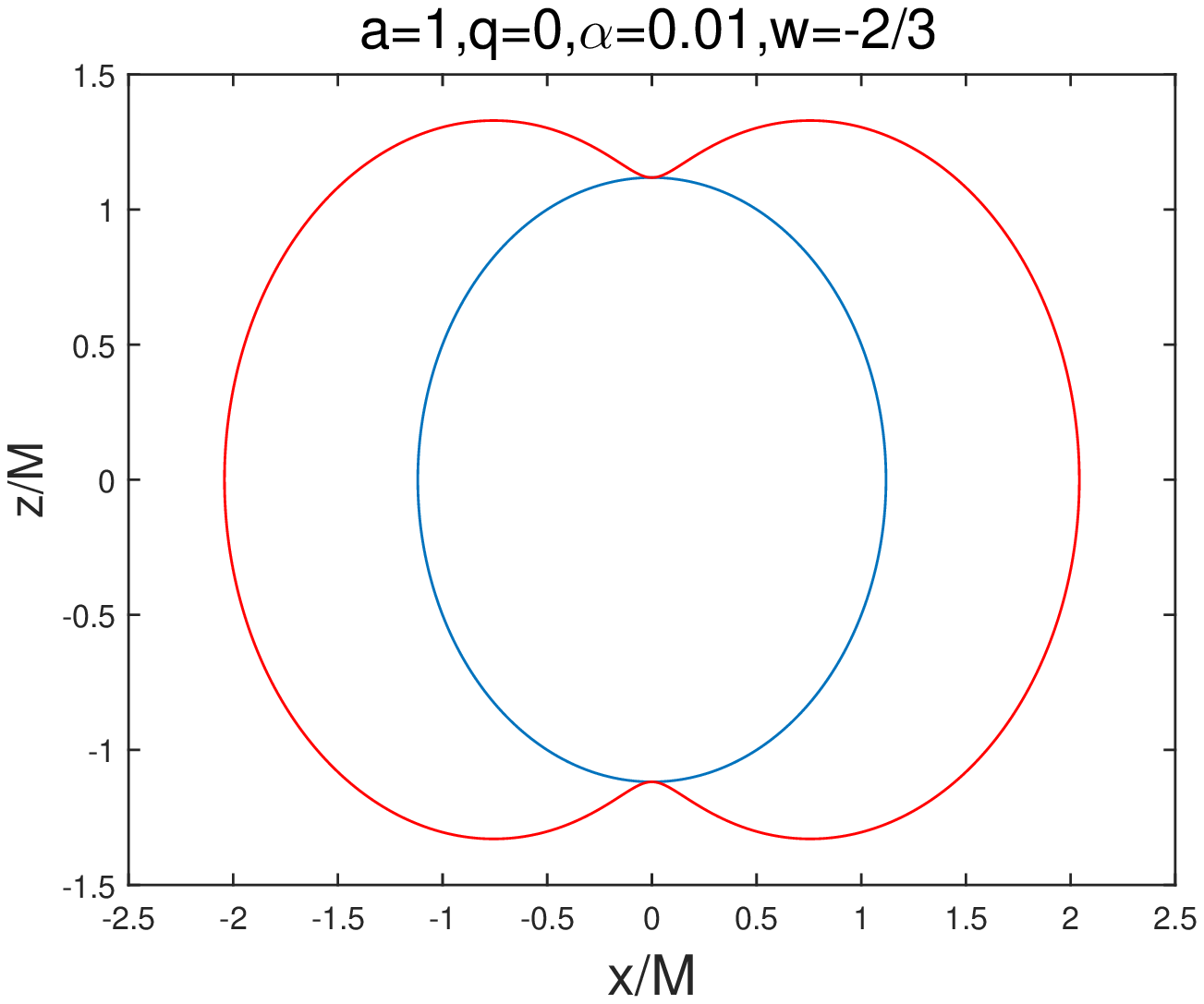}
  \includegraphics[scale=0.36]{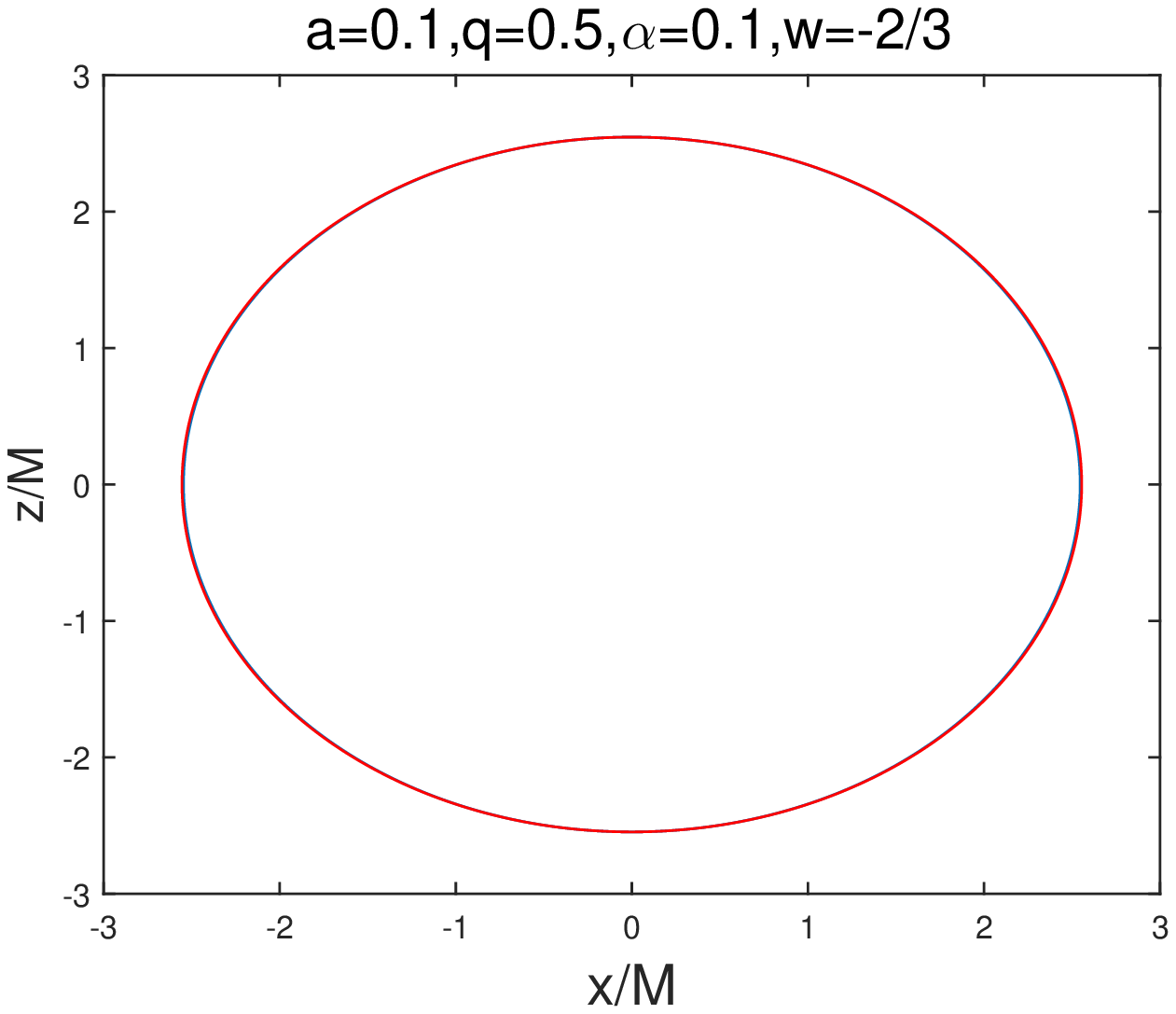}
  \includegraphics[scale=0.36]{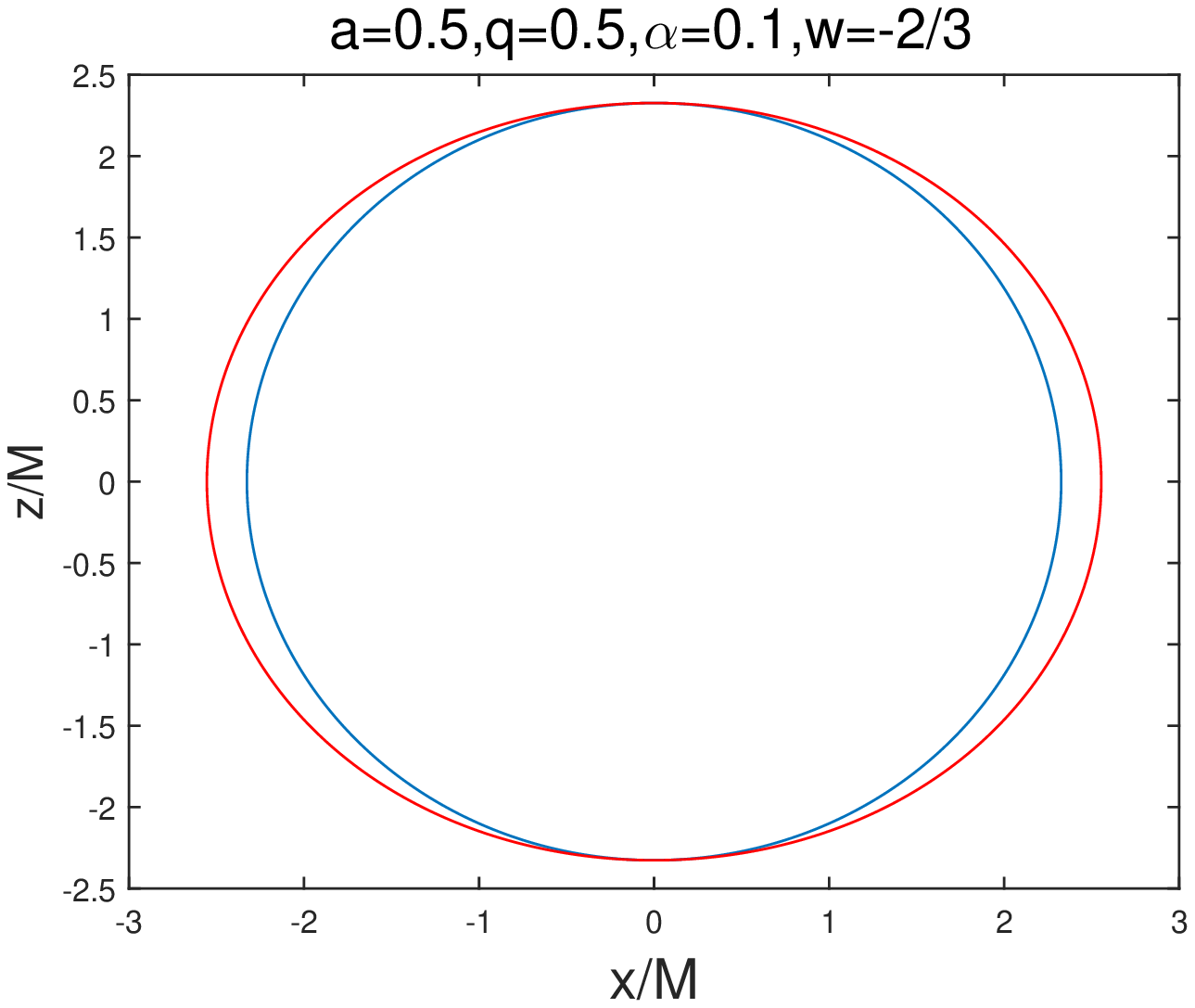}
  \includegraphics[scale=0.36]{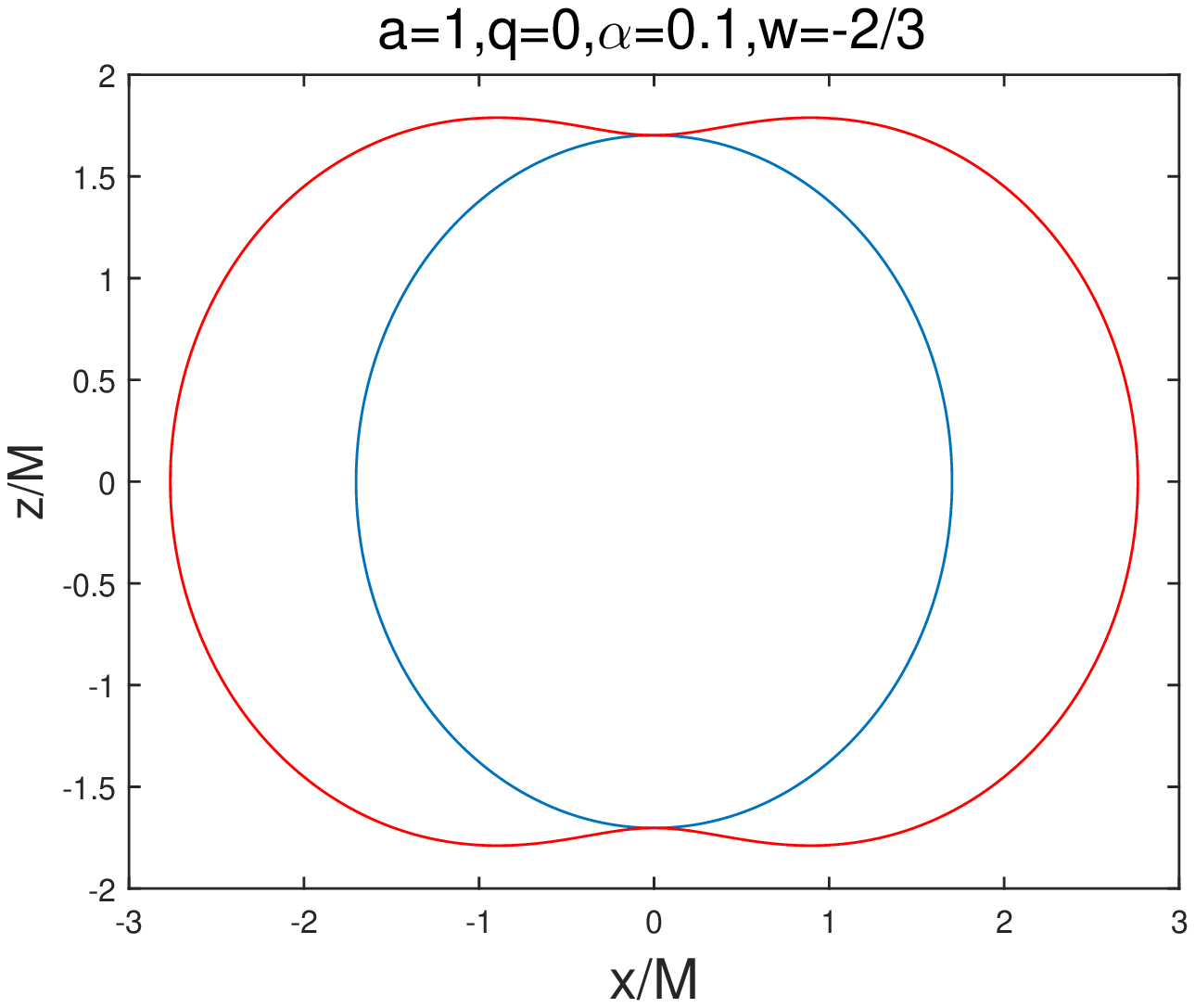}
  \includegraphics[scale=0.36]{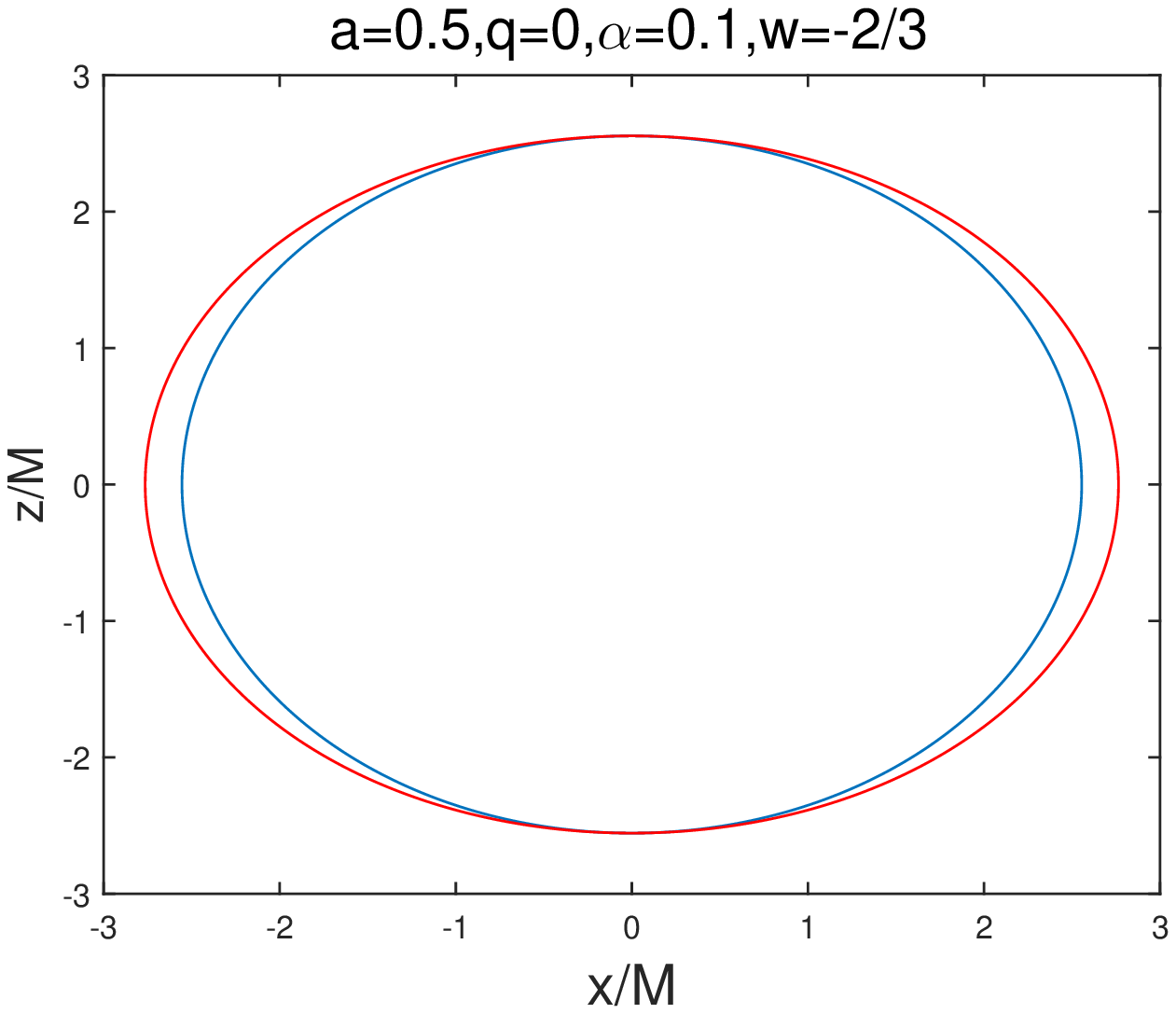}
  \includegraphics[scale=0.36]{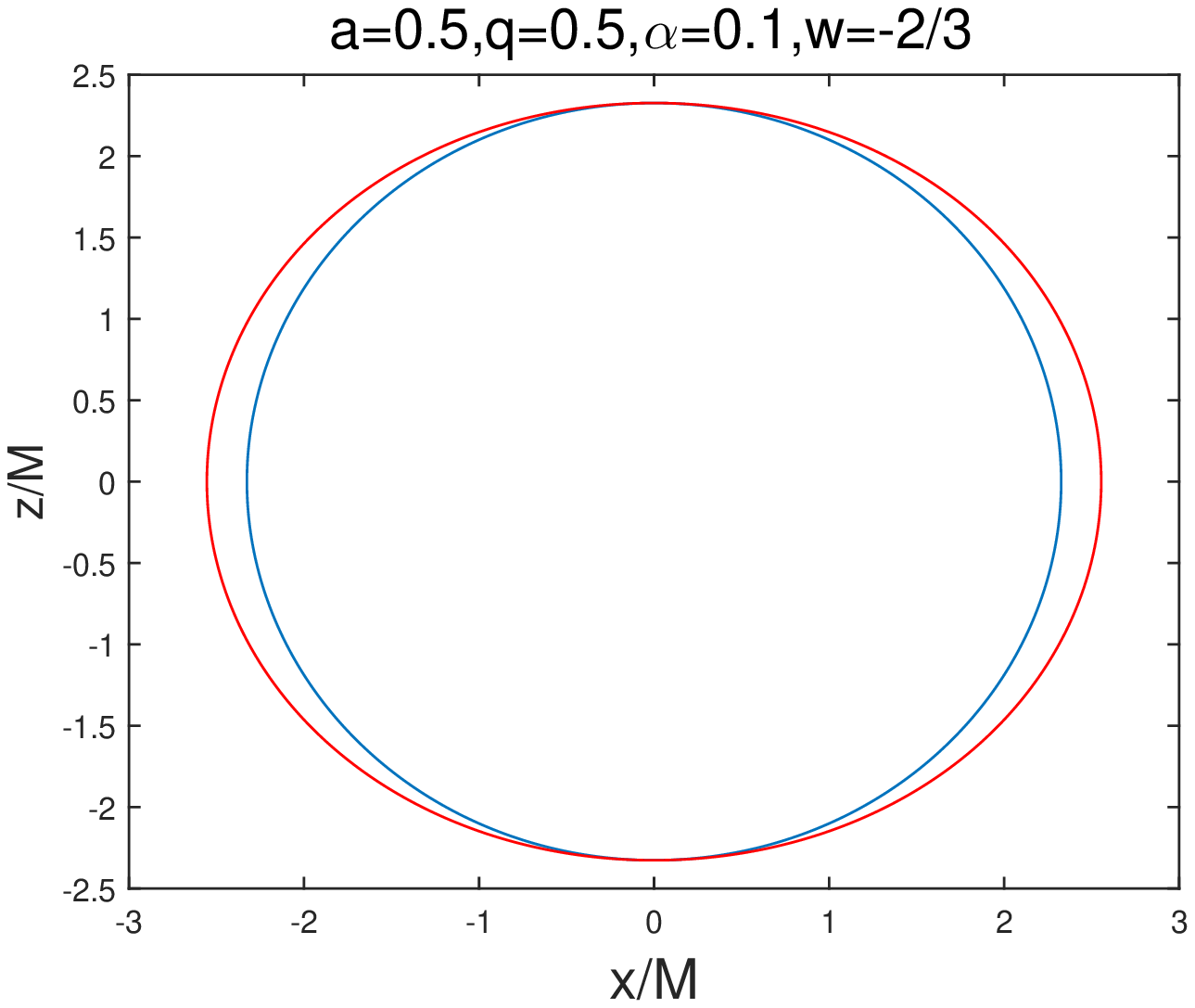}
  \includegraphics[scale=0.36]{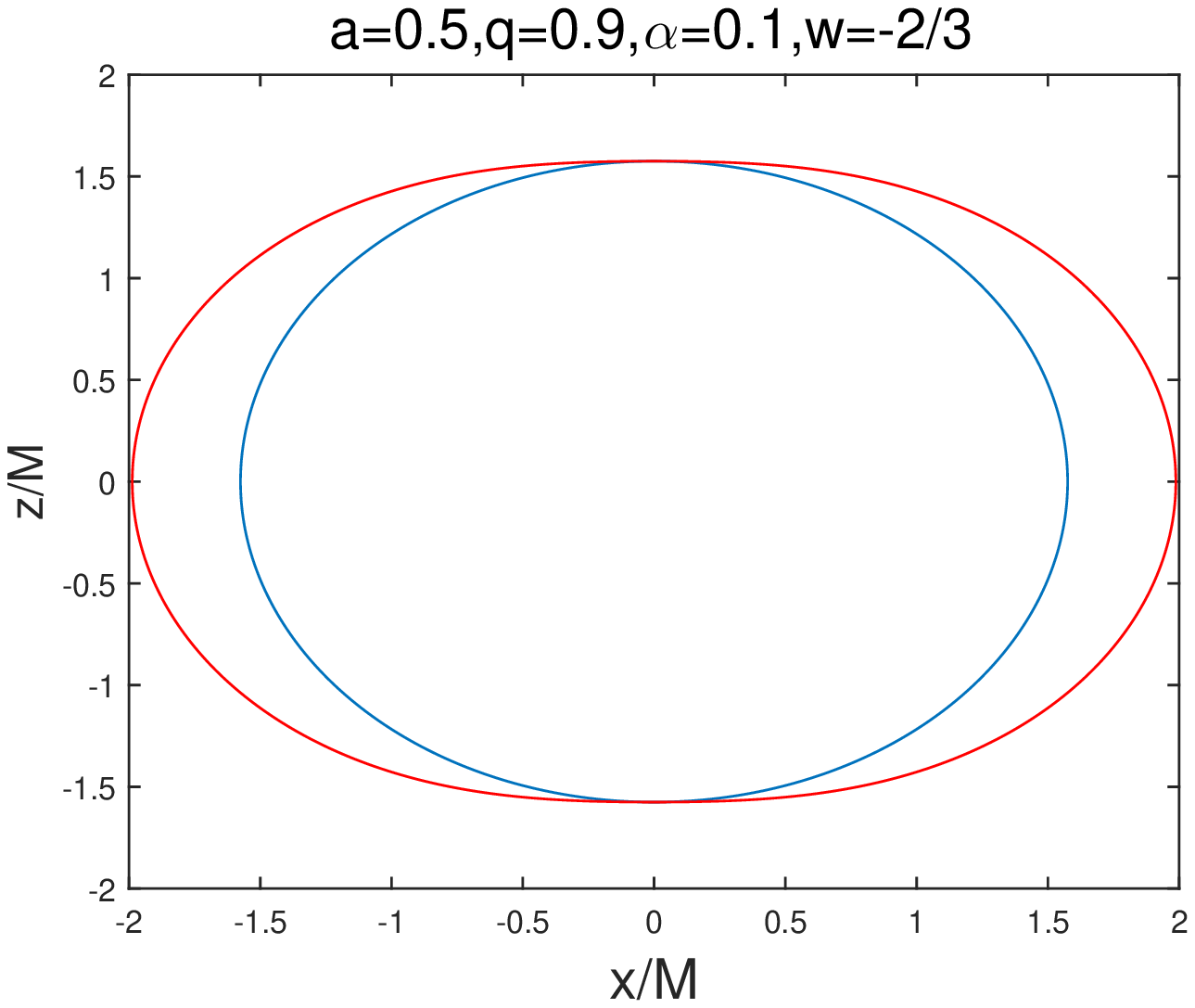}
  \caption{The shape of ergosphere of the Kerr-Newman-AdS black hole in quintessence for different $a,Q(=q),\alpha$ and $\omega=-2/3$. The blue lines represent event horizons and red lines represent stationary limit surfaces. The region between event horizon and stationary limit surface is the ergosphere. Because the cosmological constant is small, its influence can be ignored. Here $\Lambda=1.3\times10^{-56}cm^{-2}$.}
  \label{fig:2}
\end{figure}

\subsection{Stationary Limit Surfaces}
The stationary limit surfaces of the Kerr black hole have interesting properties and are defined by $g_{tt}=0$.
From the metric (\ref{16}), $g_{tt}=0$ becomes the following equation
\begin{equation}
g_{tt}=\dfrac{1}{\Sigma^{2}\Xi^{2}}(a^{2}sin^{2}\theta\Delta_{\theta}-\Delta_{r})=0.
\label{28}
\end{equation}
Following the similar equation (\ref{20}), we make this equation to become
\begin{equation}
Q^{2}+a^{2}cos^{2}\theta=-r^{2}+2Mr+\alpha r^{1-3\omega}+\dfrac{\Lambda}{3}a^{4}sin^{2}\theta cos^{2}\theta+\dfrac{\Lambda}{3}r^{2}(r^{2}+a^{2})
\label{29}
\end{equation}

There are two surfaces, e.g., out event horizon and static limit surface. They meet the poles and exist a region between horizon and static limit surface, called the ergosphere. The shape of the ergosphere is determined by the parameters $a,\omega,q,\alpha,\Lambda$ and $\theta$, and is shown in Figure 2.

\subsection{Singularities}
It's interesting to study the singularity of the black hole. By calculating the scale curvature $R$ in the metric (\ref{16}) given by$R=R^{\mu\nu\rho\sigma}R_{\mu\nu\rho\sigma}$, we can study the singularity of the black hole. The black hole is determined by $\omega$, for general $\omega$ we obtain
\begin{equation}
R=R^{\mu\nu\rho\sigma}R_{\mu\nu\rho\sigma}=\dfrac{4H(r,\theta,a,\alpha,Q^{2})}{\Sigma^{12}},
\label{35}
\end{equation}
where $H$ are polynomial function about $r,\theta$ and $a$. The function also includes $\alpha, Q, \omega$ and $\Lambda$.

We find that only $\Sigma^{2}=r^{2}+a^{2}cos^{2}\theta=0$, the real singularity exists and is given by
\begin{equation}
r=0~~~and~~~\theta=\dfrac{\pi}{2}.
\label{37}
\end{equation}
Here, we calculate the scale curvature $R$ in Boyer-Lindquist coordinates, that $\Sigma^{2}=r^{2}+a^{2}cos^{2}\theta=0$ represent a ring at the equatorial plane with the radius $a$, centered on the symmetry axis of this black Hole. It's the same with one in Kerr Black Hole (\cite{1963PhRvL..11..237K}).

\section{ROTATION VELOCITY IN THE EQUATORIAL PLANE APPLICATION TO DARK MATTER}

We derive the relation between the space-time metric components and the rotation velocity. For simplicity, we focus on the rotation motion near the equatorial plane with $\theta=\pi/2$ and $\dfrac{d\theta}{dt}=0$.

We describe the rotation curves in four-dimensional space-time formalism. The observer is in the ZAMO (zero angular momentum observers), the four-velocity satisfies the normalized condition
\begin{equation}
g_{\mu\nu}u^{\mu}u^{\nu}=-1,
\label{40}
\end{equation}
and here we consider the space-time with rotational symmetry. There are two conserved quantities as
\begin{equation}
P_{\mu}\xi^{\mu}=L, E.
\label{41}
\end{equation}
Using the expressions of $u^{\mu}$ and $u^{\nu}$, we rewrite the
normalized condition equation as
\begin{equation}
g_{tt}(\dfrac{dt}{d\tau})^{2}+2g_{t\phi}\dfrac{dt}{d\tau}\dfrac{d\phi}{d\tau}+g_{\phi\phi}(\dfrac{d\phi}{d\tau})^{2}+g_{rr}(\dfrac{dr}{d\tau})^{2}=-1.
\label{42}
\end{equation}
Using the equations (\ref{41}) and (\ref{42}), we get the following equation
\begin{equation}
-E\dfrac{dt}{d\tau}+L\dfrac{d\phi}{d\tau}+g_{rr}(\dfrac{dr}{d\tau})^{2}=-1.
\label{46}
\end{equation}
Through calculating, we obtain the equation
\begin{equation}
(\dfrac{dr}{d\tau})^{2}=-\dfrac{1}{g_{rr}}+\dfrac{g_{\phi\phi}E^{2}+2g_{t\phi}EL+g_{tt}L^{2}}{(g^{2}_{t\phi}-g_{tt}g_{\phi\phi})g_{rr}}=E^{2}-V^{2}
\label{47}
\end{equation}

\begin{figure}[htbp]
  \centering
  \includegraphics[scale=0.47]{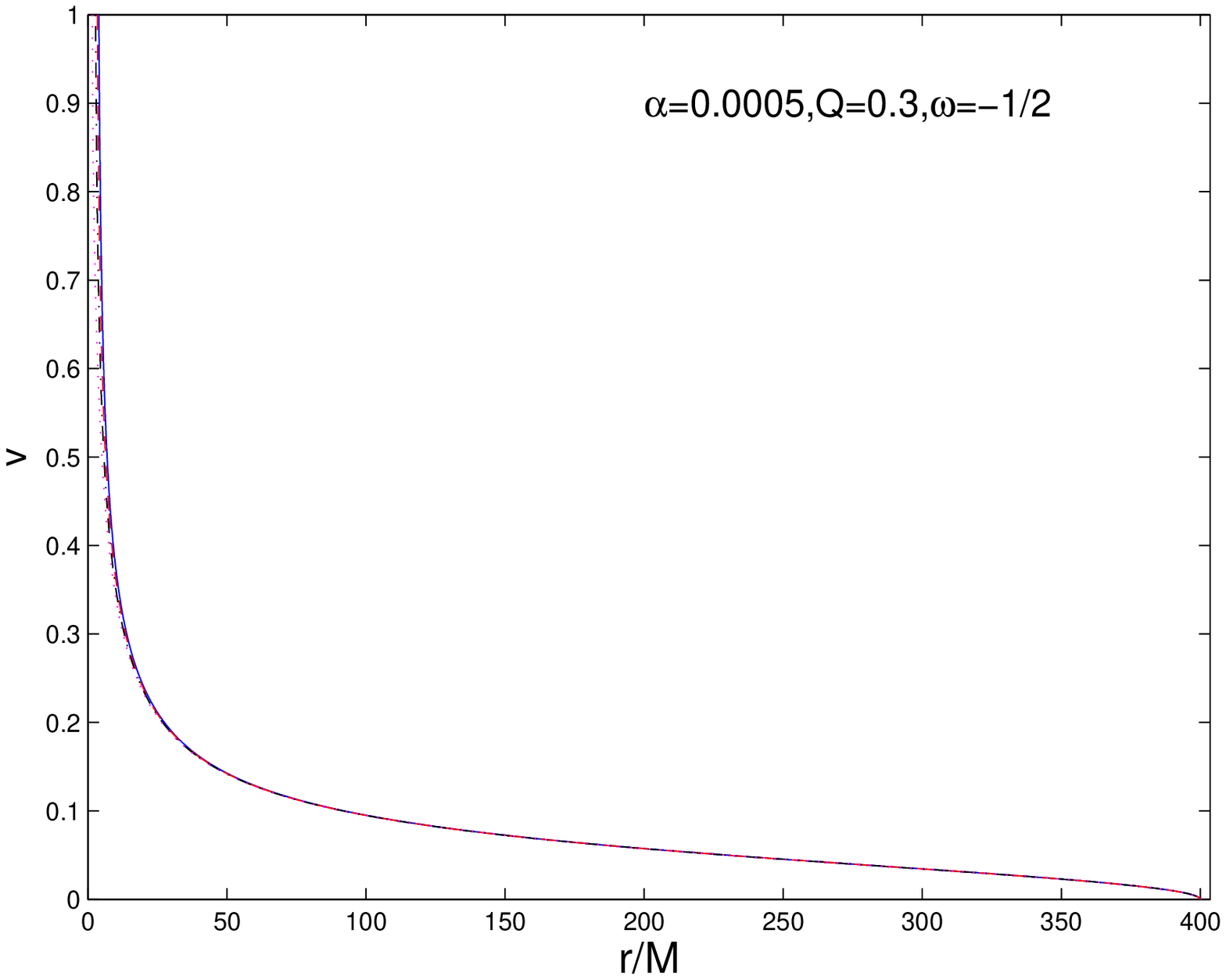}
  \includegraphics[scale=0.47]{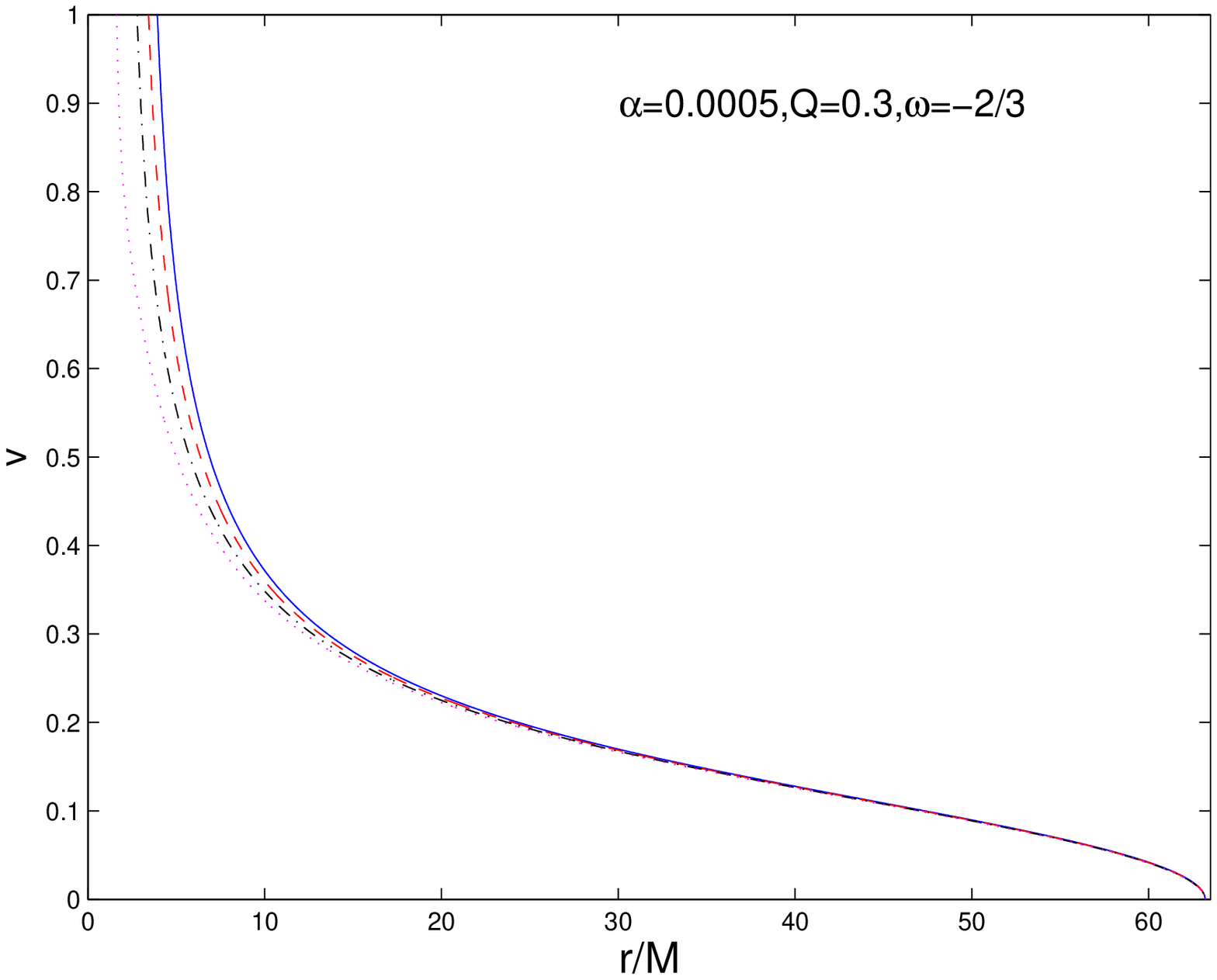}
  \includegraphics[scale=0.47]{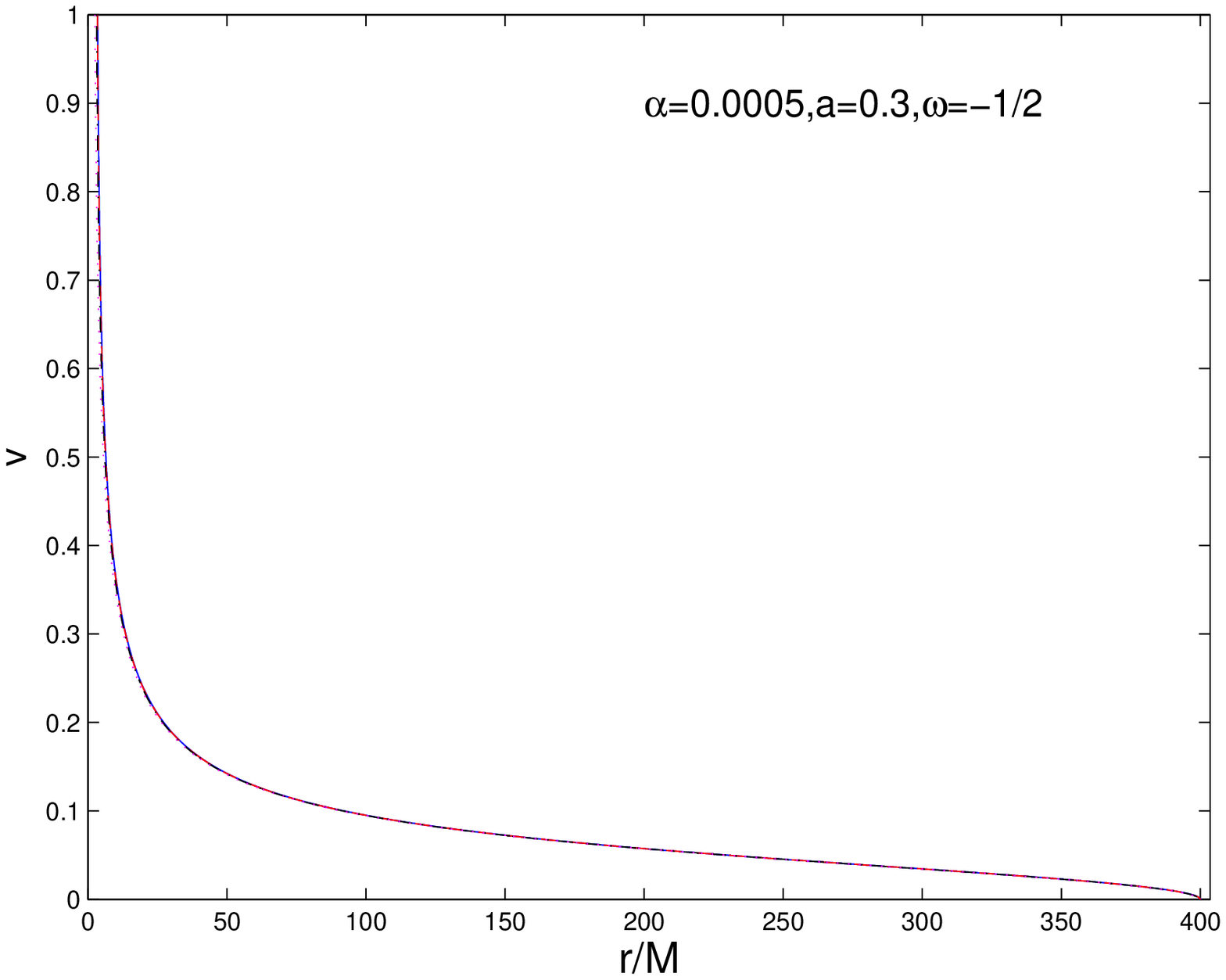}
  \includegraphics[scale=0.47]{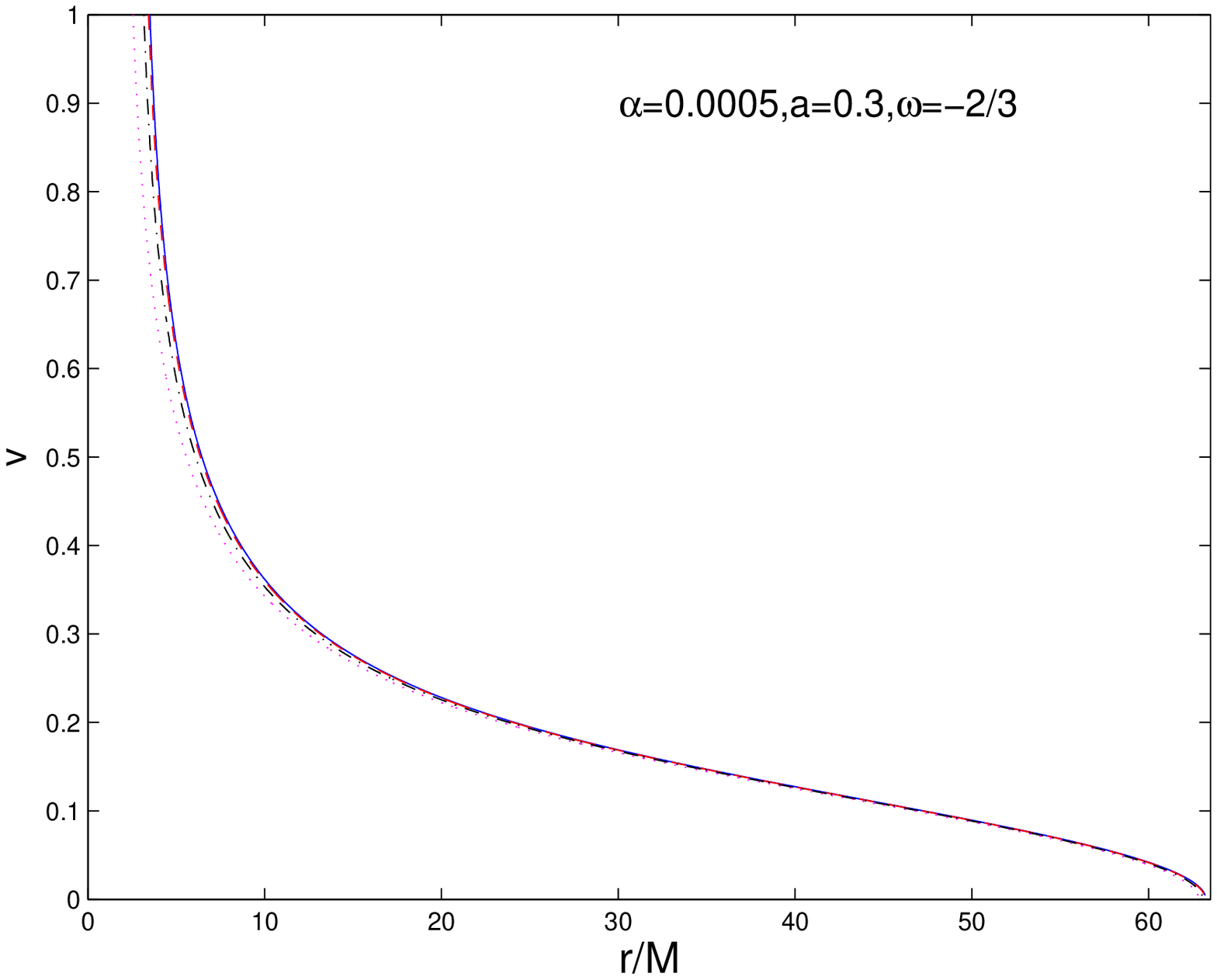}
  \caption{The behavior of rotation velocity $v$ with $r$ in the equatorial plane of the Kerr-Newman-AdS black hole in quintessential dark energy for two examples $\omega=-1/2$ and $\omega=-2/3$. The top panels show the curves for different parameter $a$: solid line $a=0$, dashed line $a=0.3$, dotdashed line $a=0.6$ and dotted line $a=0.9$. The bottom panels present the curves for different parameter $Q$: solid line $Q=0$, dashed line $Q=0.3$, dotdashed line $Q=0.6$ and dotted line $Q=0.9$. Here $\Lambda=10^{-56}cm^{-2}$ and $\alpha=0.0005$.}
  \label{fig:3}
\end{figure}

\begin{figure}[htbp]
  \centering
  \includegraphics[scale=0.55]{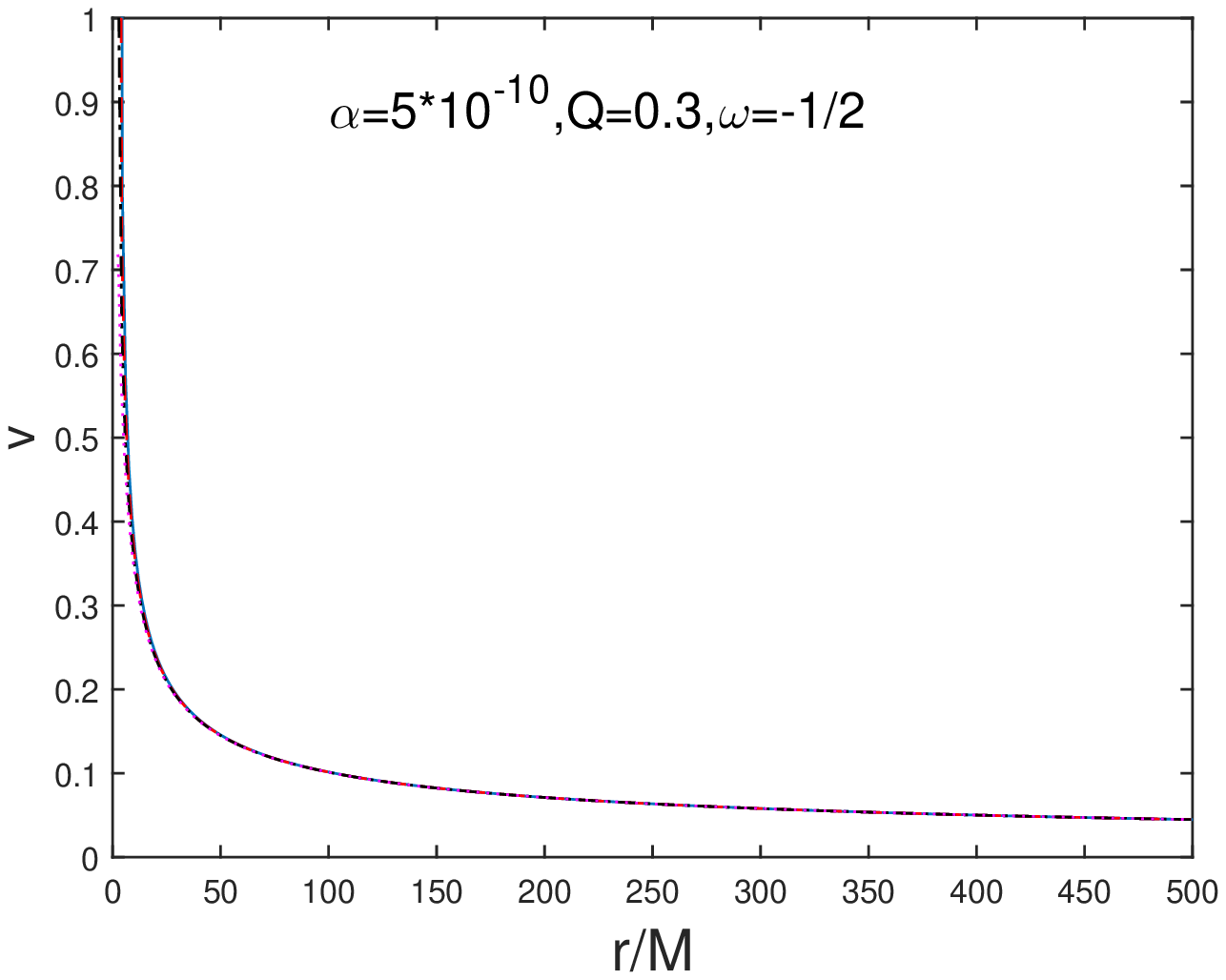}
  \includegraphics[scale=0.55]{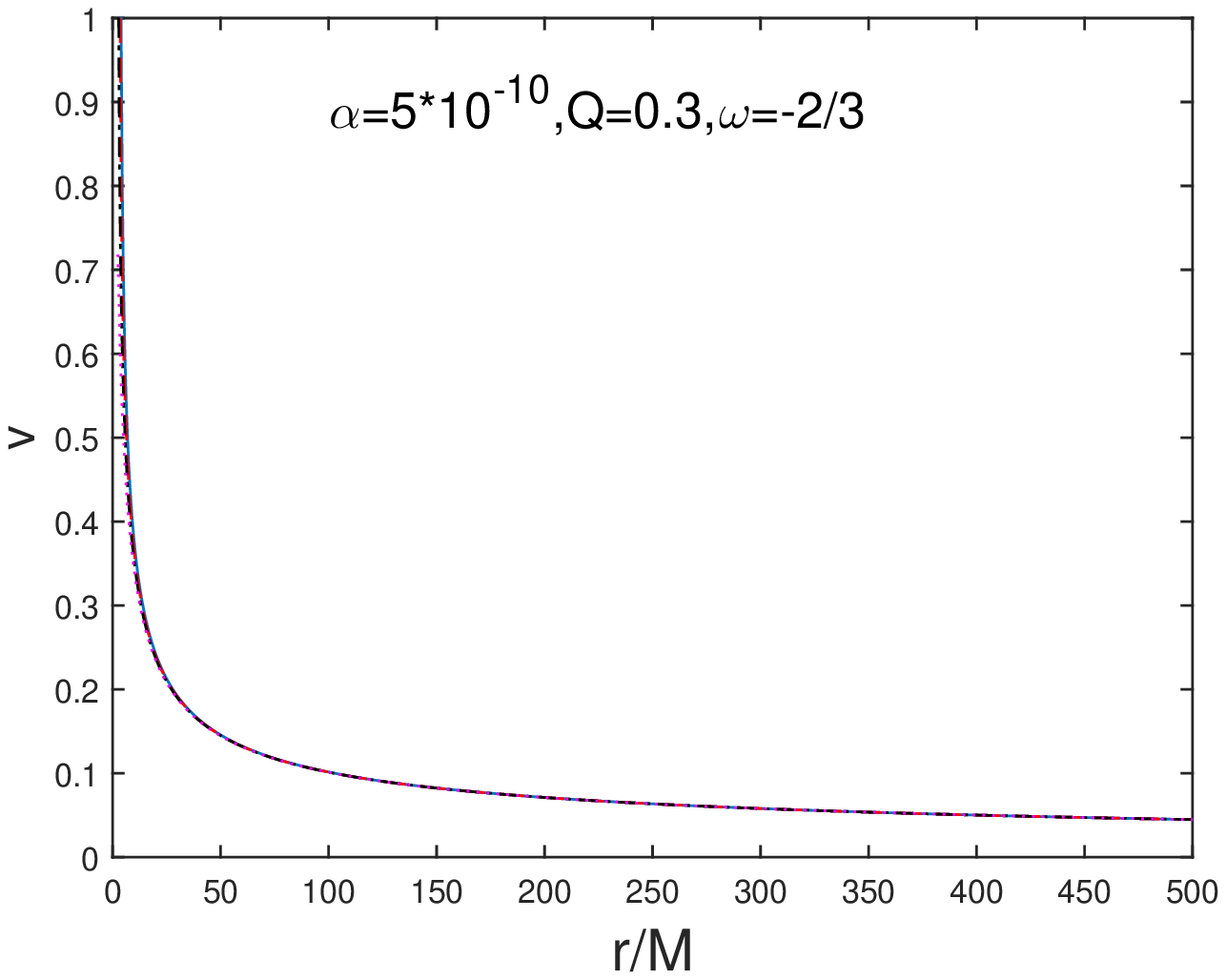}
  \includegraphics[scale=0.55]{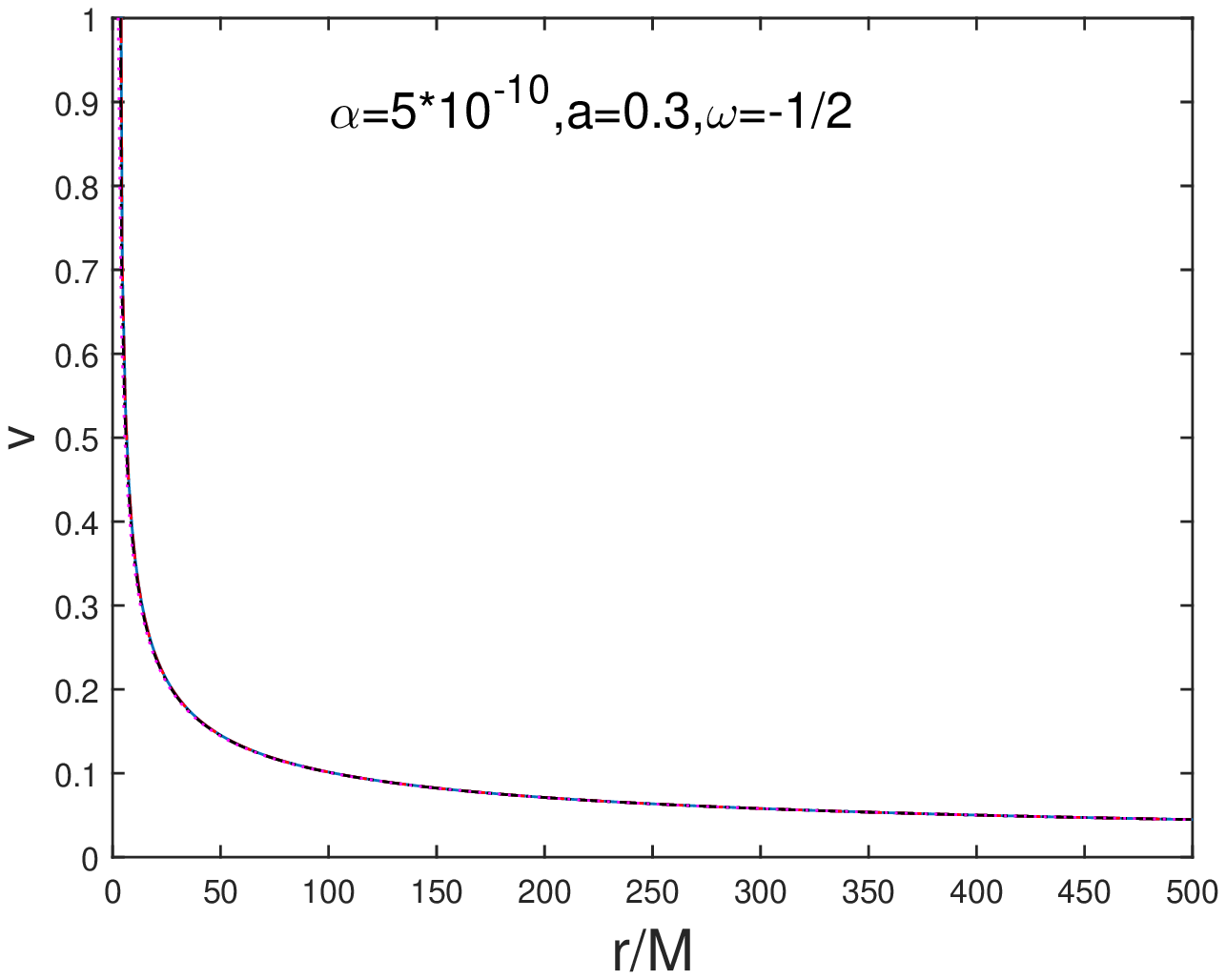}
  \includegraphics[scale=0.55]{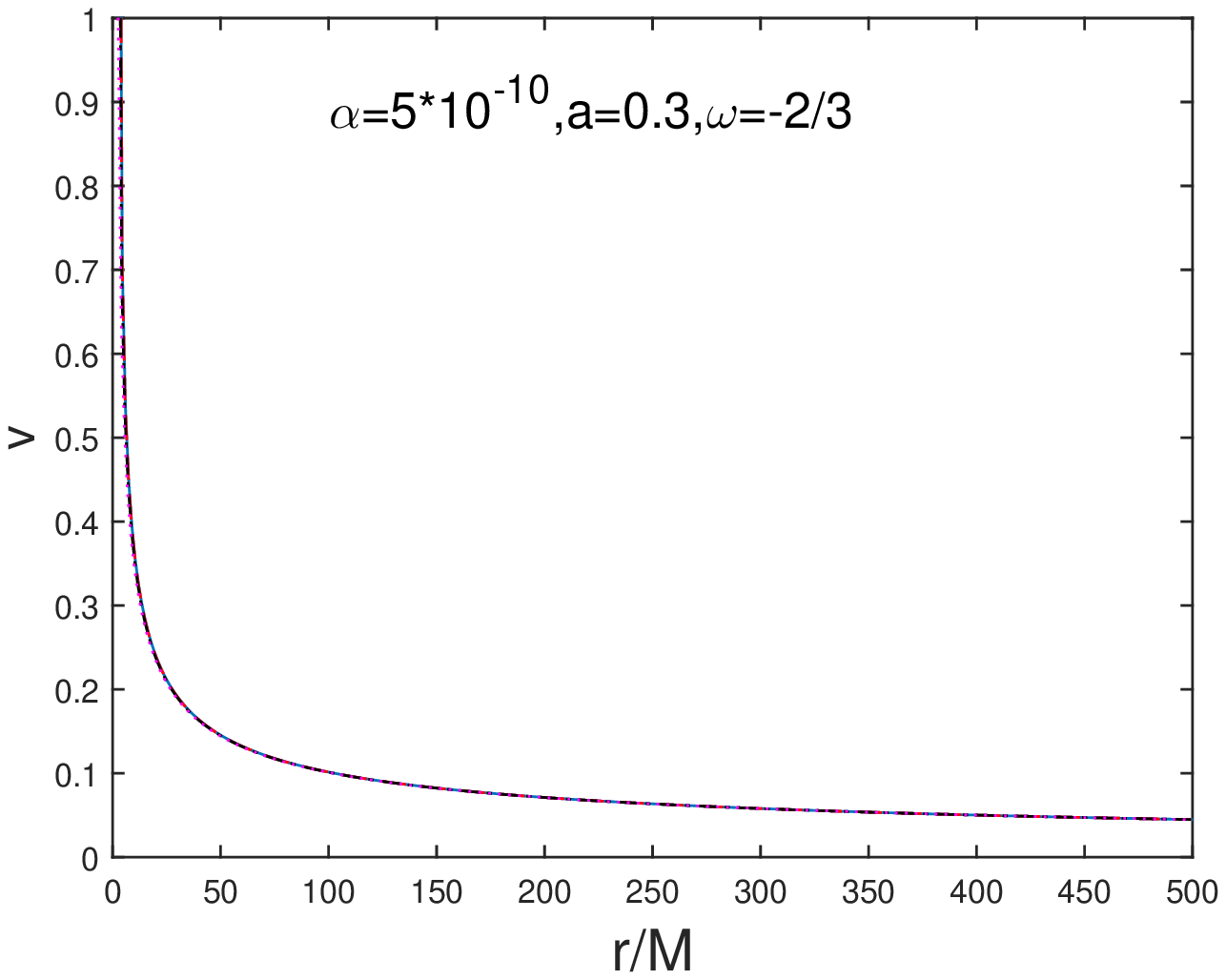}
  \caption{The behavior of rotation velocity $v$ with $r$ in the equatorial plane of the Kerr-Newman-AdS black hole in quintessential dark energy for two examples $\omega=-1/2$ and $\omega=-2/3$. The top panels show the curves for different parameter $a$: solid line $a=0$, dashed line $a=0.3$, dotdashed line $a=0.6$ and dotted line $a=0.9$. The bottom panels present the curves for different parameter $Q$: solid line $Q=0$, dashed line $Q=0.3$, dotdashed line $Q=0.6$ and dotted line $Q=0.9$. Here $\Lambda=10^{-56}cm^{-2}$ and $\alpha=5*10^{-10}$.}
  \label{fig:4}
\end{figure}

\begin{figure}[htbp]
  \centering
  \includegraphics[scale=0.55]{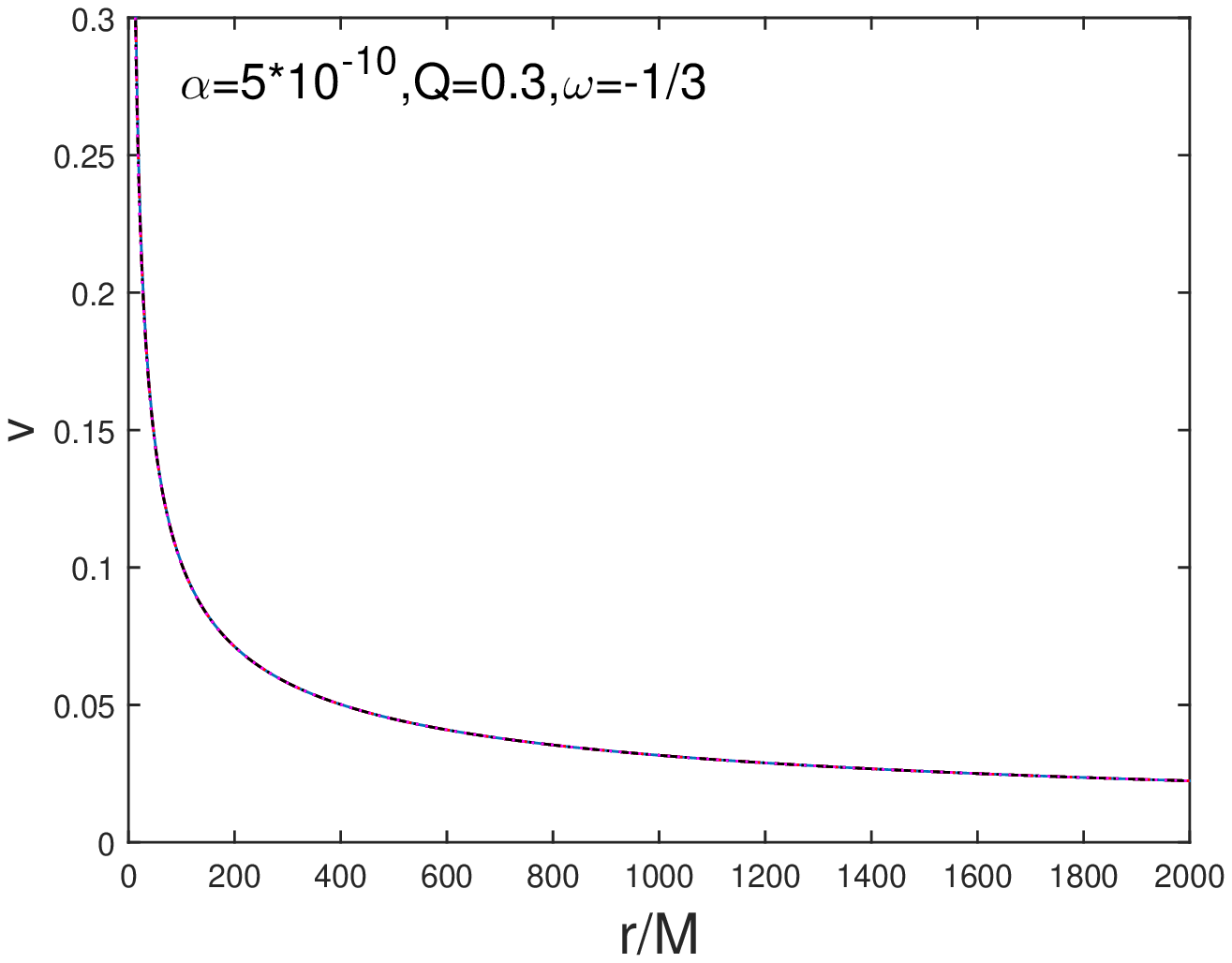}
  \includegraphics[scale=0.55]{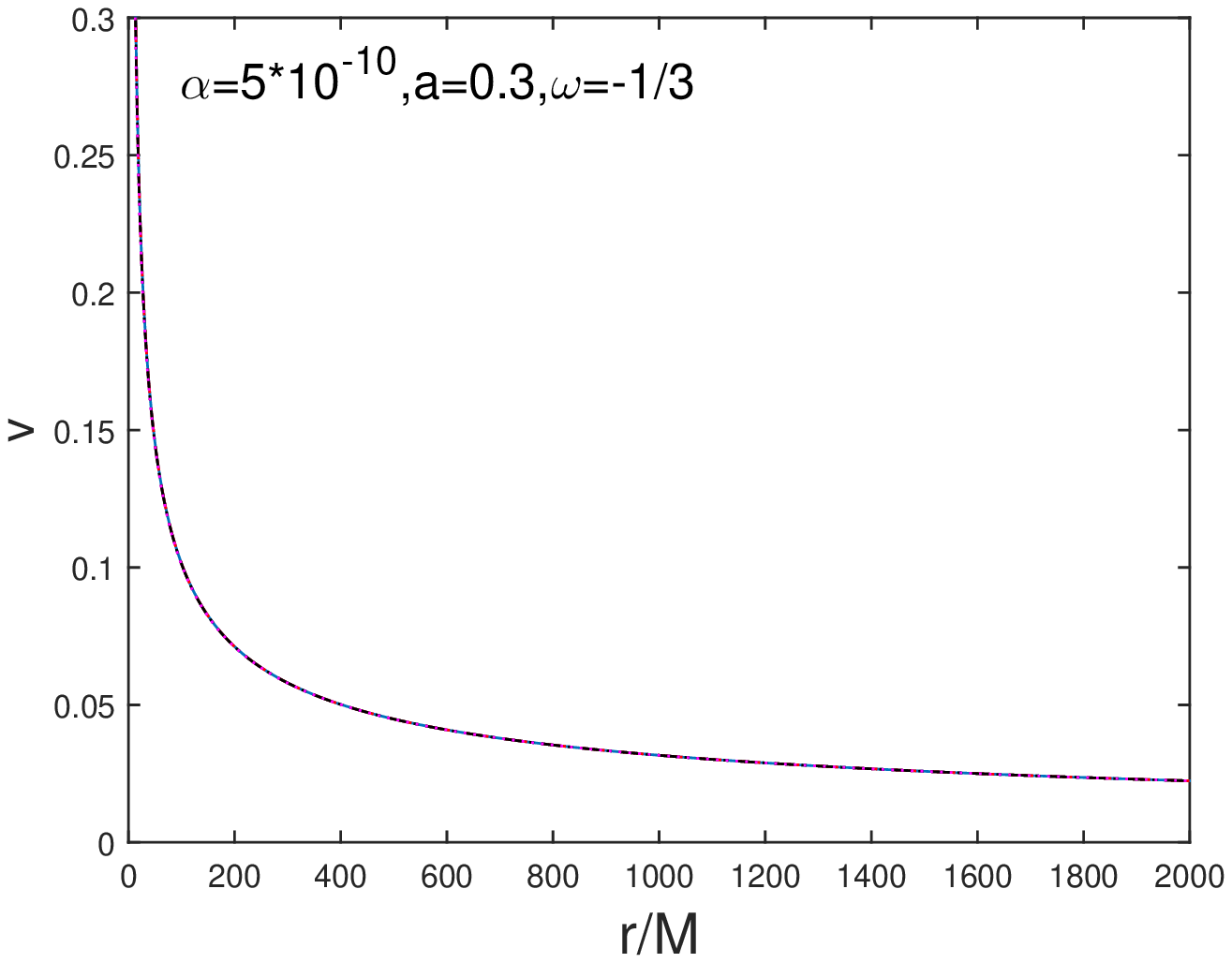}
  \caption{The behavior of rotation velocity $v$ with $r$ in the equatorial plane of the Kerr-Newman-AdS black hole in quintessential dark energy. The left picture show the curves for different parameter $a$: solid line $a=0$, dashed line $a=0.3$, dotdashed line $a=0.6$ and dotted line $a=0.9$. The right picture present the curves for different parameter $Q$: solid line $Q=0$, dashed line $Q=0.3$, dotdashed line $Q=0.6$ and dotted line $Q=0.9$. Here $\Lambda=10^{-56}cm^{-2}$ and $\omega=-1/3$.}
  \label{fig:5}
\end{figure}

The stable circular orbit satisfies two conditions
\begin{equation}
\dfrac{dr}{d\tau}=0,~~~\dfrac{dV^{2}}{dr}=0,
\label{48}
\end{equation}
Solving the equations (\ref{48}) and (\ref{47}), we obtain  (\cite{2016Ap&SS.361..269O,2013PhRvD..88d4002J})
\begin{equation}
E=\pm\dfrac{g_{tt}+g_{t\phi}\Omega_{\phi}}{\sqrt{-g_{tt}-2g_{t\phi}\Omega_{\phi}-g_{\phi\phi}\Omega^{2}_{\phi}}},~~~L=\pm\dfrac{g_{t\phi}+g_{\phi\phi}\Omega_{\phi}}{\sqrt{-g_{tt}-2g_{t\phi}\Omega_{\phi}-g_{\phi\phi}\Omega^{2}_{\phi}}},
\label{50}
\end{equation}
where the angular velocity is defined by
\begin{equation}
\Omega_{\phi}=\dfrac{-g_{t\phi,r}+\sqrt{(g_{t\phi,r})^{2}-g_{tt,r}g_{\phi\phi,r}}}{g_{\phi\phi,r}}.
\label{52}
\end{equation}

The rotation velocity for any $\omega$ is given by the following equation

\begin{equation}
v=\dfrac{L}{\sqrt{g_{\phi\phi}}}=\dfrac{1}{\sqrt{g_{\phi\phi}}}\dfrac{g_{t\phi}+g_{\phi\phi}\Omega_{\phi}}{\sqrt{-g_{tt}-2g_{t\phi}\Omega_{\phi}-g_{\phi\phi}\Omega^{2}_{\phi}}},
\label{55}
\end{equation}
where the parameter $\omega$ dominates the circular orbits. The rotation velocities on the equatorial plane are shown in Figure 3 and Figure 4 for two examples $\omega=-2/3$ and $\omega=-1/2$, they are different from those in Kerr black hole. At the same time, $\omega$ have large influence on circular orbits. When $\omega$ is close to $-1/3$ and $\alpha$ is close to $0$, the rotation velocity on the equatorial plane is more asymptotically flat. We take different charge $Q$ to draw the rotation velocities, we find that $Q$ has weak influence on the rotation velocities in the equatorial plane. Because the cosmological constant is small, its influence on rotation velocity can be ignored.

Comparing figure 3, figure 4 and figure 5, we find that when the parameter $\alpha$ is very small, the rotation velocities on the equatorial plane will be asymptotically flat in large distance $r$. Kiselev suggest that when quintessential dark energy work, the rotation curves in spiral galaxies will be asymptotically flat with distance $r$ \citep{2003gr.qc.....3031K}. In their paper, they study the rotation velocities in spherically symmetric black hole in quintessential dark energy. Here we generalize their results to Kerr-Newman-AdS black hole around by quintessential dark energy.

\section{Summary}
Using Newman-Janis algorithm, we obtain Kerr-Newman solutions in quintessential dark energy. Because Newman-Janis algorithm do not include the cosmological constant, we cannot use this method to derive Kerr-Newman-AdS solution around by quintessential dark energy. Through direct complex computation, we extend the Kerr-Newman solution to Kerr-Newman-AdS in quintessential dark energy. By analysing the horizon equation, we obtain the value of $\alpha$ for $\omega=-2/3, -1/2$. When $\Lambda=0$, we find that $\alpha\leq\sqrt{2}/5$ for $\omega=-1/2$ and  $\alpha<1/6$ for $\omega=-2/3$ which is the same with one given by \citep{2015arXiv151201498T} in quintessential dark energy, showing that the black hole charge cannot change the value of $\alpha$. When $\Lambda\neq 0$ and four horizons especially $r_{q}$ exist, we obtain the constraint equation on $\alpha$, implying that the black hole spin and cosmological constant make the maximum value of $\alpha$ to become more small. With the state parameter $\omega$ ranging from $-1$ to $-1/3$, the maximum value of $\alpha$ change $\Lambda$ to $1$. If $\omega\rightarrow -1$, $r_{q}$ arrives at $r_{c}$ and $\alpha$ is close to the cosmological constant. For all Kerr-Newman-AdS solution in quintessential dark energy, the naked singularity appears when $\Sigma^{2}=0$. Finally, we calculate the geodetic motion on equatorial plane for three situations of $\omega=-2/3, -1/2$ and $-1/3$. We find that the parameters $Q, a, \Lambda$ have small influence on rotation velocity, while the parameters $\alpha$ and $\omega$ have large influence on rotation velocity. For small value of $\alpha$, the rotation velocity on the equatorial plane is asymptotically flat and it can explain the rotation curves in spiral galaxies.

The Kerr-Newman-AdS solution around by quintessential dark energy maybe useful in astrophysics. In the future we want to study the effects of rotation and charge in a more thorough manner, and the influence of quintessential dark energy on Blandford-Znajek mechanism and black hole accretion disk.

\acknowledgments
We acknowledge the anonymous referee for a constructive report that significantly improved this paper. We acknowledge the financial support from the National Natural Science Foundation of China 11573060, 11661161010.

\end{document}